\documentclass[11pt]{article}
\pdfoutput=1

\usepackage[margin=1.0in]{geometry}
\usepackage{amsmath,amssymb,graphicx}
\usepackage{hyperref}
\usepackage{slashed}
\usepackage{wasysym}
\usepackage{hhline}
\usepackage[utf8]{inputenc}
\usepackage[greek,english]{babel}
\usepackage{subcaption} 
\usepackage{booktabs}
\usepackage{soul}
\usepackage{tikz}
\usetikzlibrary{shapes.geometric, arrows}
\usepackage[font=small,labelfont=bf]{caption}
\usepackage{cite}
\usepackage{enumerate}
\usepackage[normalem]{ulem}

\definecolor{gray}{cmyk}{0,0,0,0.05}

\newcommand{\be}{\begin{equation}}
\newcommand{\ee}{\end{equation}}
\newcommand{\bea}{\begin{eqnarray}}
\newcommand{\eea}{\end{eqnarray}}

\numberwithin{equation}{section}

\begin{document}


\begin{center}


~
\vskip 1cm

{\LARGE
\bf 
Machine Learning Classification \\ \vskip 1.5mm
of Sphalerons and Black Holes  
at the LHC }

\vskip 1.5cm

{
Aurora Singstad Grefsrud$^{(a)}$,
Trygve Buanes$^{(a)}$,
Fotis Koutroulis$^{(c)}$,
\\
\vspace{1.5mm}
Anna Lipniacka$^{(b)}$,
Rafa\l{} Mase\l{}ek$^{(c),(e)}$,
Andreas Papaefstathiou$^{(d)}$,
Kazuki Sakurai$^{(c)}$,
\\
\vspace{1.5mm}
Therese B. Sjursen$^{(a)}$
and Igor Slazyk$^{(a)}$
}

\vskip 1cm

$^{(a)}${\em
Department of Computer Science, Electrical Engineering and Mathematical Sciences, Western Norway University of Applied Sciences, \\
Postbox 7030, 5020 Bergen, Norway}
\vskip 0.2cm

$^{(b)}${\em
Department of Physics and Technology, University of Bergen, \\
Postboks 7803, N-5020 Bergen, Norway}

\vskip 0.2cm

$^{(c)}${\em
Institute of Theoretical Physics, Faculty of Physics,\\
University of Warsaw, ul.~Pasteura 5, PL-02-093 Warsaw, Poland 
}

\vskip 0.2cm

$^{(d)}${\em
Department of Physics, Kennesaw State University,\\
830 Polytechnic Lane, Marietta, GA 30060, USA
}

\vskip 0.2cm

$^{(e)}${\em
Laboratoire de Physique Subatomique et de Cosmologie (LPSC), Université Grenoble-Alpes,
CNRS/IN2P3, 53 Avenue des Martyrs, F-38026 Grenoble, France}

\end{center}

\vskip 1.3cm

\begin{abstract}
In models with large extra dimensions, 
``miniature'' black holes (BHs) might be produced in high-energy proton-proton collisions at the Large Hadron Collider (LHC).
In the semi-classical regime, those BHs thermally decay, giving rise to large-multiplicity final states with jets and leptons.  
On the other hand, similar final states are also expected in the production of electroweak sphaleron/instanton-induced processes.  
We investigate whether one can discriminate these scenarios when BH or sphaleron-like events are observed
in the LHC using machine learning (ML) methods. 
Classification among several BH scenarios with different numbers of extra dimensions and the minimal BH masses is also examined. 
In this study we consider three ML models: XGBoost algorithms with (1) high- and (2) low-level inputs, and (3) a Residual Convolutional Neural Network.
In the latter case, the low-level detector information is converted into an input format of three-layer binned event images, where the value of each bin corresponds to the energy deposited in various detector subsystems.
We demonstrate that only a small number of detected events are sufficient to effectively discriminate between the sphaleron and BH processes. 
Separation between BH scenarios with different minimal masses is possible with an order of 10 events passing the preselection. 
A sufficient number of events could be observed in combined Run-2 and -3 data, if the production cross section is not much smaller than the present limit $\sim 0.1$ fb.
We find, however, that a large number of events is needed to discriminate between BH hypotheses with the same minimal BH mass, but different numbers of extra dimensions.

\end{abstract}


\vskip 1.5cm

\newpage
\section{Introduction}
\label{sec:intro}

The Standard Model (SM) of particle physics has been extremely successful in describing particle interactions below the TeV scale. 
However, the SM has several theoretical issues, such as the hierarchy problem, as well as phenomenological problems like the inability to account for the existence of dark matter and the observed asymmetry between matter and antimatter in the Universe. These problems suggest that a more fundamental theory underlying the SM must exist, and may emerge at the energy scale around, or higher, than the TeV scale.  

An attractive solution to the hierarchy problem is a scenario with \emph{Large Extra Dimensions} (LEDs)  \cite{Arkani-Hamed:1998jmv, Antoniadis:1998ig, Arkani-Hamed:1998sfv}. This model postulates that all matter and gauge fields, except for gravity, are restricted to live on a $(3+1)$-dimensional hypersurface, called a 3-brane, which is embedded in the higher dimensional spacetime. Assuming all extra dimensions orthogonal to the 3-brane are compactified, the traditional Planck scale $M_P \sim 10^{18}$ GeV is understood as an effective scale, derived from the fundamental higher-dimensional Planck scale $M_*$ with the relation:
\be
M_*^{2+n} \sim \frac{M_P^2}{R^n}
\,,
\ee
where we have assumed the existence of $n$ extra dimensions with the common size $R$. If the size of the extra dimensions is much larger than the traditional Planck length, $R \gg 1/M_P$, the fundamental Planck scale $M_*$ can be much smaller than the usual Planck scale $M_P$. In particular, the hierarchy problem is solved if $M_*$ is around the TeV scale \cite{Arkani-Hamed:1998jmv, Antoniadis:1998ig, Arkani-Hamed:1998sfv}.

This scenario has a striking implication for new physics searches at high-energy colliders, such as the Large Hadron Collider (LHC) at CERN. For the fundamental Planck scale around or smaller than the TeV scale, $M_* \lesssim {\cal O}(1)$ TeV, high-energy colliders can offer trans-Planckian particle collisions with $\sqrt{\hat s} > M_*$. 
In this situation, the gravitational interaction becomes very strong and the colliding particles may collapse into a black hole (BH) \cite{Banks:1999gd, Dimopoulos:2001hw, Eardley:2002re,Yoshino:2002br, Yoshino:2002tx, Ida:2002ez,Harris:2003db, Harris:2004xt}. 
In the semi-classical regime, where the BH mass is significantly larger than $M_*$,
such a ``miniature" BH, once formed, will quickly ``evaporate'' through Hawking radiation \cite{Hawking:1975vcx}.\footnote{ 
As opposed to the semi-classical regime, production and decay of quantum BHs (QBHs) have been discussed \cite{Giddings:2001bu,Meade:2007sz},
and collider searches have been carried out.
Those searches are based on the assumption that QBHs decay exclusively into two high-energy SM particles.
The current limit depends on the decay channels:
$\lesssim 1$ fb for dijet \cite{QBH_dijet} and $\lesssim 0.1$ fb for lepton+jet \cite{QBH_lj} and
charged lepton flavour violating final channels (${\rm QBH} \to e\mu, e \tau, \mu \tau$) \cite{QBH_LFV}.
If these limits are considered seriously, there is little chance that the LHC can observe semi-classical BHs.
However, the dynamics of QBHs are much less known than those of their semi-classical counterparts, and in this paper, we assume the cases where the current limit on QBH is not applicable. 
}
The spectrum of emitted particles can be understood as the thermal radiation characterized by the Hawking temperature $T_H$ of the BH at a given stage of the evaporation \cite{Myers:1986un, Giddings:2001bu}.
Those outgoing particles, however, must travel through a strong gravitational potential created by the BH, and the spectrum is distorted from the blackbody profile for an observer located infinitely far from the BH \cite{DeWitt:1975ys,Page:1976df}. The deviation from the blackbody spectrum is described by the so-called ``greybody factor'', which carries the information of extra dimensions \cite{Giddings:2001bu}.

The last observation poses an interesting question to collider physics: ``can we discriminate one extra dimension scenario from another
with the BH signature at a collider, if observed?'', or more concretely, for example, ``can we identify the number of extra dimensions by analysing BH events?''.
Since the greybody factor depends on the parameters of extra dimensions in a subtle way, answering these questions is non-trivial. 

There even exists a similar collider signature that has a completely different origin, due to electroweak (EW) sphaleron/instanton-induced processes \cite{Manton:1983nd,Klinkhamer:1984di, Belavin:1975fg, Affleck:1980mp}.
These are non-perturbative processes within the SM, which go over (or penetrate via quantum tunneling) a potential barrier separating two distinctive EW vacua: the Higgs and EW gauge field configurations in these vacua are characterised by different values of the topological winding number, $N_{\rm CS}$ (the Chern-Simons number).
The EW sphaleron plays a crucial role in many proposed scenarios of baryogenesis (see, e.g., Ref.\ \cite{Rubakov:1996vz} for a review). 
The theoretical prediction for the production cross section of these non-perturbative processes at high-energy colliders suffers from large uncertainties, and whether such processes are observable in the foreseeable future is still under debate~\cite{Ringwald:1989ee, Espinosa:1989qn, Tinyakov:1992dr, Khlebnikov:1990ue, Mueller:1991fa, Khoze:1990bm, Diakonov:1993ur, Balitsky:1993xc, Ringwald:2002sw, Bezrukov:2003er, Tye:2015tva, Ellis:2016ast, Ellis:2016dgb, 
Brooijmans:2016lfv,
Funakubo:2016xgd, Tye:2017hfv, Ringwald:2018gpv, Cerdeno:2018dqk,Papaefstathiou:2019djz,Jaeckel:2022osh}. Although the cross section is largely unknown, the final state of the process is anticipated to possess certain characteristic features. Firstly, the minimum height of the potential barrier, $E_{\rm sph}$, is known {to} good accuracy, $E_{\rm sph} \simeq 9.1$ TeV \cite{Manton:1983nd,Klinkhamer:1984di}. Therefore, one expects that those processes can occur only at very high energies, at least with $\sqrt{\hat s} \gtrsim E_{\rm sph}$. Secondly, since the anomaly connects the change of $N_{\rm CS}$, and the change of the fermion number that couples to the $SU(2)_L$ gauge bosons, \emph{all} left-handed fermions, ($3$ quarks $+$ $1$ lepton) $\times$ (3 generations), of the SM must be involved in the interaction \cite{tHooft:1976snw, tHooft:1976rip}. Therefore, `consuming' two light-flavor quarks in the initial state, the final state must contain at least seven anti-quarks and three anti-leptons (plus some EW bosons). Such a high-energy and high-multiplicity final state resembles the BH events mentioned. It is therefore a non-trivial task to discriminate the EW sphaleron/instanton-induced processes from the miniature BH signature in the LED scenario. 

A traditional way of tackling the above questions is to analyse various event variables built out of reconstructed objects.
In particular, in Ref.\ \cite{CMS:2018ozv}, the CMS collaboration has performed an analysis to search for semi-classical BHs and EW sphalerons, resulting in the cross section limits of the order of 0.1 fb.\footnote{They set the model-independent limit  
$\lesssim 0.1$ fb 
on the cross section times acceptance 
in the signal regions with $N \gtrsim 5$ and $S_T \gtrsim 7$ TeV 
or $N \gtrsim 10$ and $S_T \gtrsim 4$ TeV
(see the definitions of $N$ and $S_T$ in the next section).
In the sphaleron scenario, this corresponds to the prefactor $\lesssim 0.02$ with the threshold energy $E^{\rm sph}_{\rm thr} \gtrsim 9$ TeV.
For the BH scenario with a certain assumption (non-rotating, $n=6$, the fundamental Planck scale $M_* = 4$ TeV) and the Hoop conjecture \cite{Thorne:1972ji}, this limit translates into the minimal BH mass $M_{\rm min} \gtrsim 10$ TeV. 
For more details, see Ref.\ \cite{CMS:2018ozv}.}
The CMS event selection is based on the reconstructed high-$p_T$ objects, with $p_T > 70$ GeV, including jets, isolated electrons, muons and photons.
Although this approach appears to be powerful for searches, it has several drawbacks when applied to discrimination problems in high-multiplicity final states.
For example, we will show later that isolated leptons are potentially a powerful discriminator between the BH and EW sphaleron scenarios.
However, charged leptons arise only in some fraction of signal events.  Moreover, those leptons are often rejected by the isolation criterion in the busy environment of a high-multiplicity event. 
As a result, the majority of signal events contain no reconstructed leptons.
This is a crucial problem for our case since the expected signal cross sections are low ($\lesssim 0.1$ fb).
Ref.\ \cite{Dimopoulos:2001hw} proposed a method to measure the number of extra dimensions by fitting the Hawking temperature formula 
with the electron/photon spectrum 
in BH events.
Given the current cross section limit, however,
this method is no longer available 
as it requires large statistics. 
Ideally, the optimal discrimination method should use 
as much information as possible in the high multiplicity event data, rather than looking at a particular object.  
Finding the optimal analysis method or constructing a good discrimination variable, however, becomes increasingly difficult when the number of final state objects increases.

%


The identification of the number of extra dimensions, and the discrimination between the BH and EW sphaleron/instanton events, is essentially a classification problem, which machine learning (ML) methods have been proven to be exceptionally good at.
Ref.\ \cite{Schichtel:2019hfn} studied a way to look for EW sphaleron events in cosmic ray showers with ML.
The application of ML to collider physics has also been an active research field in the last decade. Recent development includes the application of deep learning, i.e.,\ neural networks (NNs) to, for example, triggering, background estimation, jet tagging, and event classification (see, e.g., Ref.\ \cite{hepmllivingreview} for a comprehensive list of references). More recently, deep learning methods 
utilising low-level whole-event data
(e.g.,\ the energy deposits of the entire calorimeter) have been investigated \cite{Aurisano:2016jvx, Bhimji:2017qvb, Andrews2019}. These studies represented entire collider events as images, and processed them using Convolutional Neural Networks (CNNs) \cite{CNN1}. They have demonstrated that such methods can exceed the sensitivity of standard approaches with high-level inputs, such as the four-momenta of reconstructed objects.
Classifying events using low-level inputs has an obvious advantage in our problem, since the object reconstruction process is largely omitted, and most detector information is kept and used in the analysis. It should be noted, however, that this approach has certain limitations in high pile-up conditions, when the detector is swamped by energy deposits that do not belong to the main hard interaction of interest. Reconstructions of jets pointing to the main vertex of interest, and rejecting the deposits not belonging to them, typically serve as pile-up mitigation. 
In what follows, we assume that pile-up mitigation can be performed, and we do not consider energy deposits coming from pile-up interactions. Furthermore, we assume that the SM backgrounds can be suppressed with the CMS-inspired selections \cite{CMS:2018ozv}, which are applied to our data.

In this paper, we study the discrimination of EW sphaleron and five different BH scenarios, using three different ML methods. The aim is to see if we can separate these scenarios with a reasonable number of events that can be collected in the on-going and future LHC runs. 
Three ML methods are examined.
The first two are based on the state-of-the-art XGBoost library, which constructs a boosted decision tree model.
The low-level tracking and calorimeter data are used as inputs in the first method,
whereas the second method employs tabular reconstructed object data. 
The third method uses a Residual Neural Network (ResNet) model based on a CNN architecture. Here, the low-level detector information is converted into three-layer event images with a resolution of $50 \times 50$ bins and bin values corresponding to the energy deposits in the electromagnetic and hadronic calorimeters and $p_T$ of tracks observed in the tracking system.  
We proceed to estimate the expected $p$-values for each hypothesis, for a given number of observed events in the signal region, originating from each of the possible scenarios. 

The rest of the paper is organised as follows.
In section \ref{sim} our simulation setup and event selection are explained.
The six model hypotheses used in this study are also given in this section.
The distributions of several observables are studied in section \ref{kin}.
The three ML models used in this study 
are laid out in section \ref{sec:ML}.
Our main result, the comparison between 
different ML models and the exclusion $p$-values 
of the hypothesis test, is shown and discussed in section \ref{sec:results}. 
Section \ref{sec:concl} is devoted to conclusions.

\section{Monte Carlo simulation and event selection}
\label{sim}

\begin{table}
    \begin{center}
        \begin{tabular}{ c | c c c c} 
            \hline
            Hypothesis  & $n$  & $M_{\rm min}$ & $E^{\rm sph}_{\rm thr}$ & $\epsilon_{\rm sig}$\\
            \hline
            \hline
            SPH\_9 & --- & --- & 9\,TeV & 0.18 \\
            BH\_n4\_M8 & 4 & 8\,TeV & --- & 0.17\\ 
            BH\_n2\_M10 & 2 & 10\,TeV & --- & 0.55 \\
            BH\_n4\_M10 & 4 & 10\,TeV & --- & 0.49\\ 
            BH\_n6\_M10 & 6 & 10\,TeV & --- & 0.43\\ 
            BH\_n4\_M12 & 4 & 12\,TeV & --- & 0.65\\ 
            \hline 
        \end{tabular}
    \caption{List of the six model hypotheses used in this study.
    The first row is the sphaleron scenario with the 9 TeV threshold energy, $E^{\rm sph}_{\rm thr}$.
    The other five rows correspond to the BH scenarios with different number of extra dimensions $n$ and the minimal BH masses $M_{\rm min}$.
    The signal efficiencies $\epsilon_{\rm sig}$ for the signal region ($N \ge 5$, $S_T \ge 7$ TeV)
    are shown in the last column.}
        \label{hypotheses}
    \end{center}
\end{table}

To study the classification of high-multiplicity events, we first consider semi-classical BH events.
For simplicity and to align with the CMS analysis \cite{CMS:2018ozv}, we focus on non-rotating BHs and fix the fundamental Planck scale at $M_* = 4$ TeV throughout our analysis.
Within this assumption, the two most important parameters are the number of extra dimensions, $n$,
and the minimal BH mass, $M_{\rm min}$, which we vary.
The minimal BH mass is defined 
as the partonic threshold energy, where the semi-classical BH events are turned on.
Such a phenomenon is expected under the Hoop conjecture.\footnote{
As opposed to this case,
there is a scenario in string theory, where
semi-classical BH-like events turn on smoothly as a function of the partonic collision energy 
\cite{Veneziano:2004er}.
We do not consider such a scenario.
}
We treat the production cross section as an independent parameter, rather than relying on specific assumptions to connect it with other parameters.  
The observed number of events is therefore not used for the classification input. 
In this treatment, 
the model-dependent CMS bounds, $M_{\rm min} \gtrsim 10$ TeV, \cite{CMS:2018ozv}
is also not applicable.

The BH events are generated at parton level using the {\tt BlackMax} event generator~\cite{BlackMax}.
{\tt BlackMax} allows for the choice of the number of extra dimensions, $n$, and the minimal BH mass, $M_{\rm min}$.
In this study, we examine five different BH hypotheses with number of extra dimensions set to $n=2$, 4 and 6 and minimal BH masses to $M_{\rm min} = 8$, 10, and 12\,TeV. 
In the semi-classical regime, the BH loses its mass via Hawking radiation.
However, when the BH mass reaches $M_*$, the semi-classical approximation breaks.
The detailed mechanism of the final burst of the BH evaporation is not known.
The {\tt BlackMax} takes the minimal approach.
Once the BH mass reaches $M_*$, the BH decays 
into the minimal number of SM particles, which conserves all unbroken charges, such as the total electric charge, colour and four-momentum. 
The back reaction from this burst process to the BH geometry is not considered.  
Altering these assumptions might affect our conclusion quantitatively.

The EW sphaleron/instanton-induced events are generated using {\tt Herwig 7}~\cite{Bahr:2008pv, Gieseke:2011na, Arnold:2012fq, Bellm:2013hwb, Bellm:2015jjp, Bellm:2017bvx, Bellm:2019zci} with the instanton addon library \cite{Papaefstathiou2019}. 
We fix the threshold energy of the sphaleron production at $E_{\rm thr}^{\rm sph} = 9$ TeV.
Both sphaleron and black hole events are simulated assuming proton-proton collisions at $\sqrt{s}=13~\mathrm{TeV}$.
It has been argued that sphalerons may be dominantly produced in association with multiple EW bosons \cite{Ringwald:1989ee,Khoze:1990bm}.
This argument is, however, based on an approximation in a low energy regime, $E \ll E_{\rm thr}^{\rm sph}$.
We do not know whether this conclusion still holds at 
$E \sim E_{\rm thr}^{\rm sph}$.\footnote{In fact, the CMS analysis \cite{CMS:2018ozv} is based on sphaleron events without the boson production.}
We found that the sphaleron events are most similar to the BH events when the boson production is switched off (see the first plot in Fig.\ \ref{jet}).
If the multi-boson production is included,
sphaleron events would have much higher jet multiplicities than BH events 
and the discrimination between them would be trivial. 
We therefore choose the zero boson option to study the most interesting case for the discrimination problem.
For the parton shower and hadronisation, we employ
{\tt Herwig 7} for both BH and sphaleron events. 
The detector response and object reconstruction are simulated with the fast multipurpose detector response simulation framework {\tt Delphes 3} \cite{Delphes}, assuming the conditions and geometry of the ATLAS detector. This paper is a proof-of-concept study and does not include any real ATLAS data or ATLAS simulations. Instead, we use computationally inexpensive open-source tools that are good approximations of, 
not openly available, ATLAS internal simulations.
Although we do not expect a large effect from the detector modelling, a detailed study using the full detector simulation may be necessary for the real-world application.

The six model hypotheses studied in this paper are listed in Table \ref{hypotheses}.
In both BH and sphaleron scenarios, we treat the production cross sections as unknown parameters, since we wish to focus on discrimination with kinematics only, while keeping in mind, however, that the current experimental limit on the sphaleron and BH cross sections is $\sigma \lesssim 0.1$ fb \cite{CMS:2018ozv, Ringwald:2018gpv} with $E_{\rm thr}^{\rm sph}, M_{\rm min} \sim 8$-12 TeV.\footnote{ One might wonder whether one can treat the signal cross section as an independent parameter from the partonic threshold energy, $M_{\rm thr} = E_{\rm thr}^{\rm sph}$ or $M_{\rm min}$.  Such a treatment is indeed possible because the partonic cross section is 
phenomenologically modeled as $\hat \sigma(\hat s) = \hat \sigma_0 \Theta( \sqrt{\hat s} - M_{\rm thr})$
\cite{CMS:2018ozv, Papaefstathiou:2019djz},
where $\Theta(x)$ is the Heaviside step function.
The hadronic cross section, obtained by integrating $\hat \sigma(\hat s)$ with the parton distribution functions, still depends on unknown parameter $\sigma_0$, even after fixing $M_{\rm thr}$.}

Before delving into the classification problem, we would like to highlight that the SM background is under control.
In particular, in Ref.\ \cite{CMS:2018ozv}, the CMS collaboration has performed an analysis to search for semi-classical BHs and EW sphalerons and has demonstrated that the SM background, dominantly QCD multijet events, can be heavily suppressed with an appropriate event selection. In the CMS analysis, the event selection is based on the two reconstructed quantities, $N$ and $S_T$. The variable $N$ is defined as the number of high-$p_T$ objects in an event, with $p_T > 70$ GeV, including jets with $|\eta| < 5$, isolated electrons and photons with $|\eta| < 2.5$, and isolated muons with $|\eta| < 2.4$. In the following, we call the reconstructed objects satisfying the above criteria, {\it signal objects}. For example, signal electrons comprise of the isolated electrons with $p_T > 70$ GeV and $|\eta| < 2.5$. 
The $S_T$ is the scalar sum of the missing transverse energy $p_T^{\rm miss}$ and the magnitude of transverse momenta of all $N$ signal objects,
$S_T \equiv |p_T^{\rm miss}| + \sum_{i=1}^N |p_T^i|$.
The CMS study has shown that the SM background is reduced to $\sim 0.1$ event at 36 fb$^{-1}$ (13 TeV) by demanding 
$S_T \ge 7$ TeV and $N \ge 5$.
On the other hand, 17-65 \% of the signal events remain after this event selection, as shown in Table \ref{hypotheses}.
In the following analysis, we apply essentially the same event selection. The only difference is that our signal objects are defined with a tighter pseudorapidity cut, $|\eta| < 2.4$.\footnote{We have checked the change of the signal efficiency by this modification is less than 1\%.}  
Since the SM background can be safely neglected by this selection, we do not consider it further in our analysis.
Assuming the signal cross section is just below the current limit 
and the event selection efficiency is $\sim 50$ \%,
we expect around $30$ signal events after the cut at
the LHC run-3 (13.6 {\rm TeV}, 300 $\rm fb^{-1}$),
while the number of expected backgrounds is $\sim 1$ or smaller since our selection cuts are tighter than the CMS ones.\footnote{Here we took into account the fact that
by the energy increase from $13$ TeV to $13.6$ TeV,
the signal cross section is roughly doubled for $E_{\rm thr}^{\rm sph}, M_{\rm min} \sim 9$ TeV \cite{Ellis:2016ast}, while the background cross section does not change significantly.}  
At the High-Luminosity LHC (HL-LHC), with an integrated luminosity of $3$ ab,
one has to even tighten the selection cuts
to suppress the SM background below 1.
This is necessary to justify our analysis where 
the SM background is not included in the simulation. 
After such a tighter selection cut, one can still expect the observed signal events of ${\cal O}(100)$ or more,
especially if the HL-LHC is operated with the collision energy of 14 TeV. 
Although we envisage the collision energy slightly higher than 13 TeV, our simulation is performed with 13 TeV throughout, for simplicity.  
We, however, do not expect significant changes in the event kinematics and our qualitative result is valid for collision energies between 13 and 14 TeV.
Since we do not expect a large number of observed signal events, our goal is to find a model discrimination method applicable to low signal statistics, ${\cal O}(30) - {\cal O}(100)$.

\section{Kinematical distributions}
\label{kin}

\begin{figure}
     \centering
         \centering
         \includegraphics[scale=0.25]{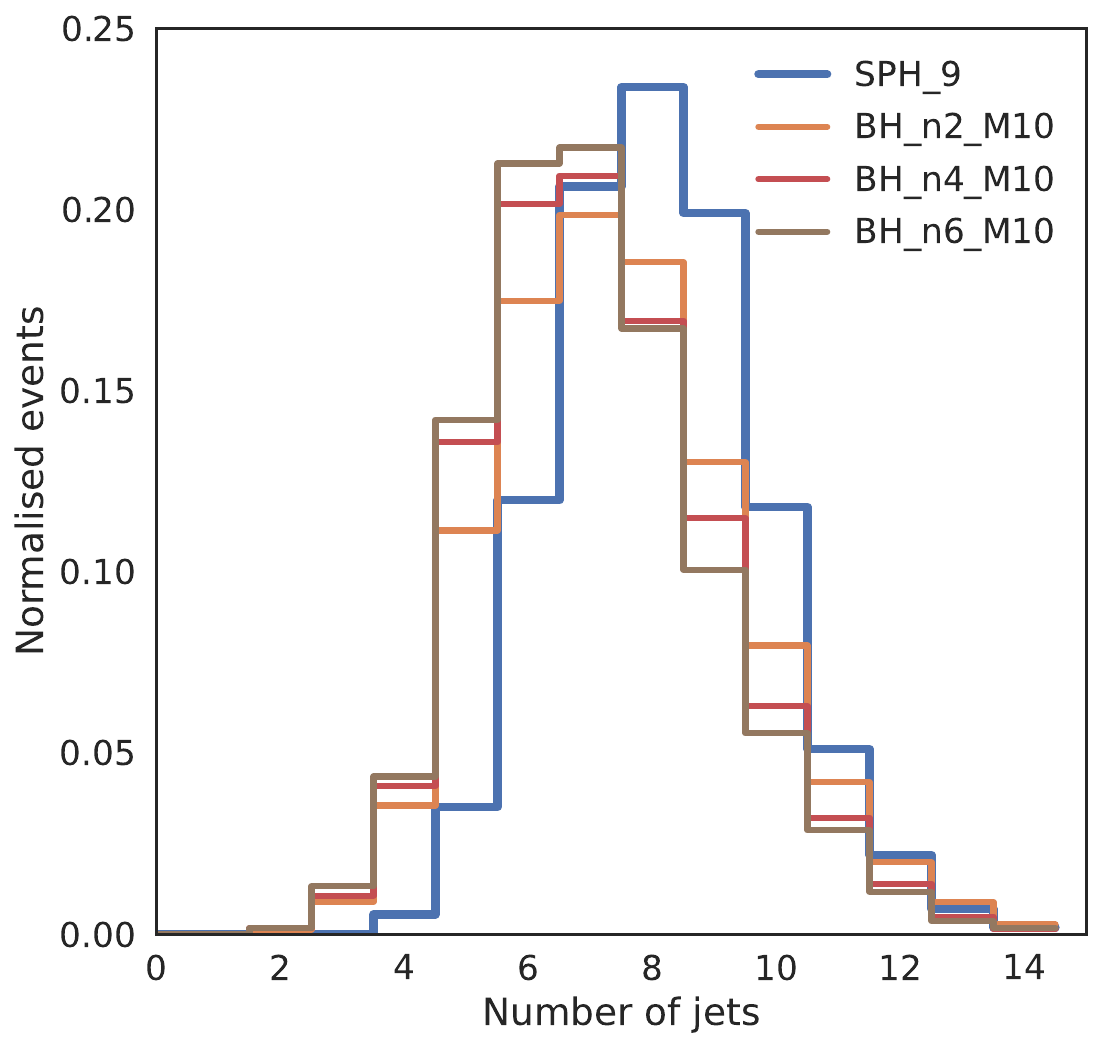}         
         \includegraphics[scale=0.25]{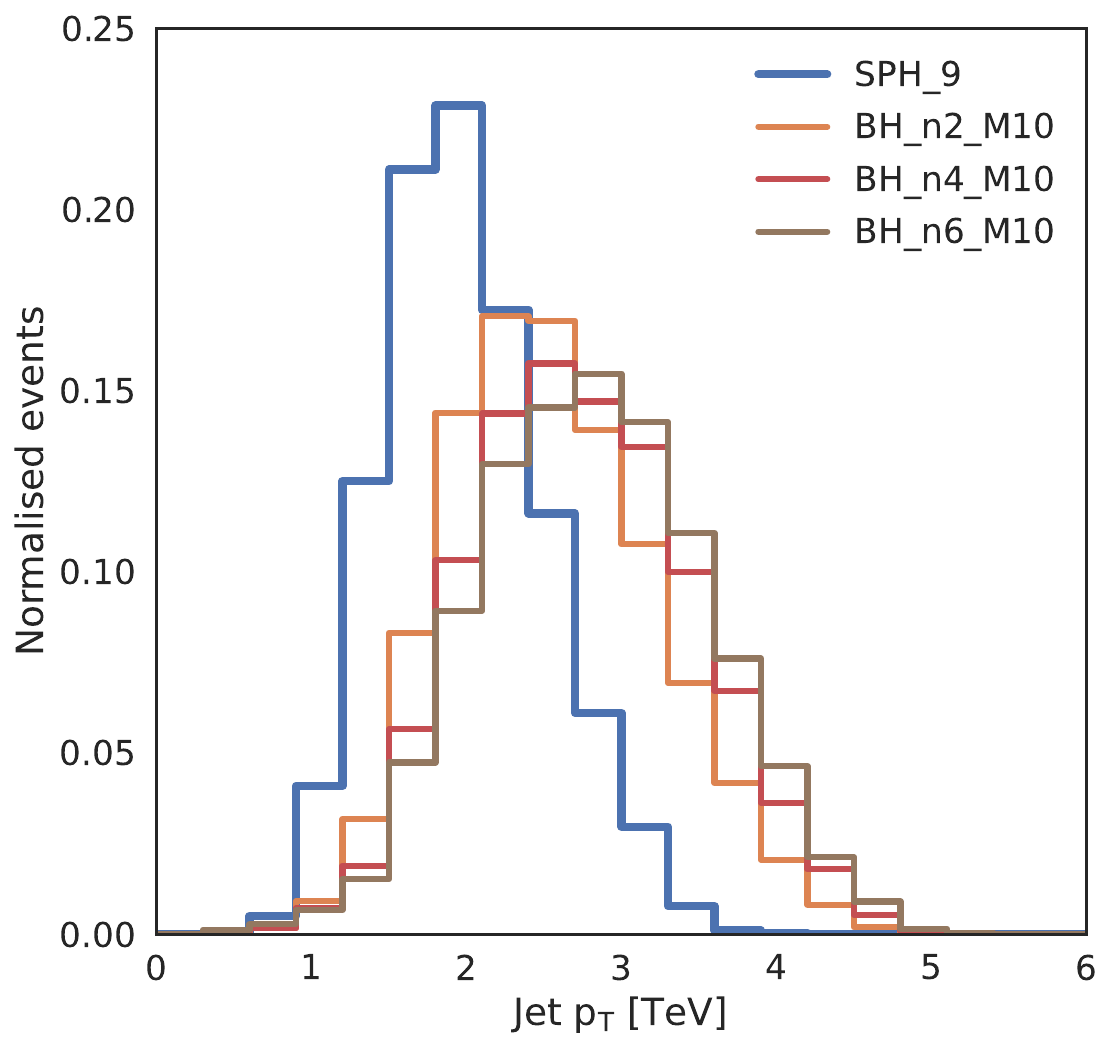}
         \includegraphics[scale=0.25]{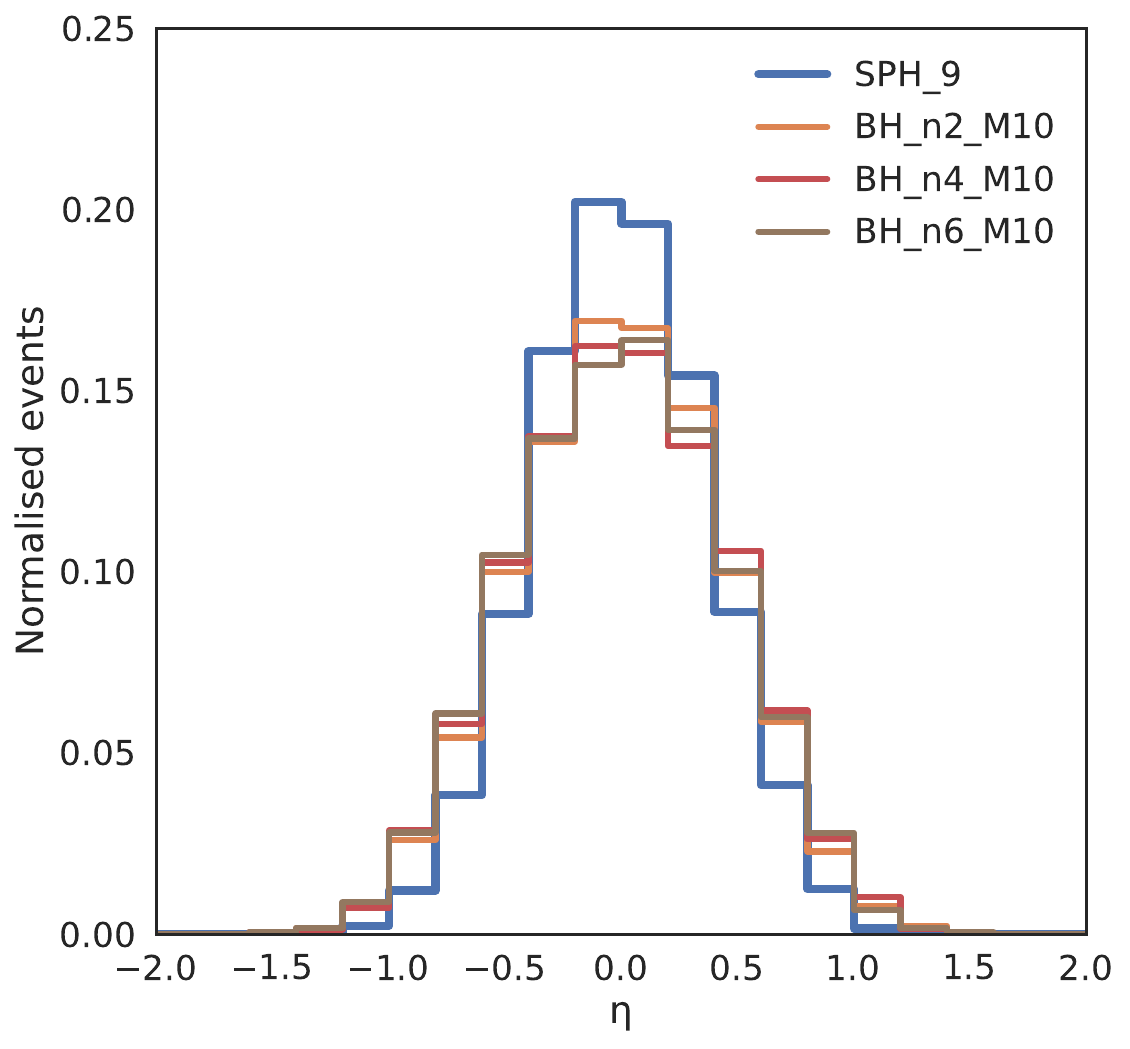}
         \includegraphics[scale=0.25]{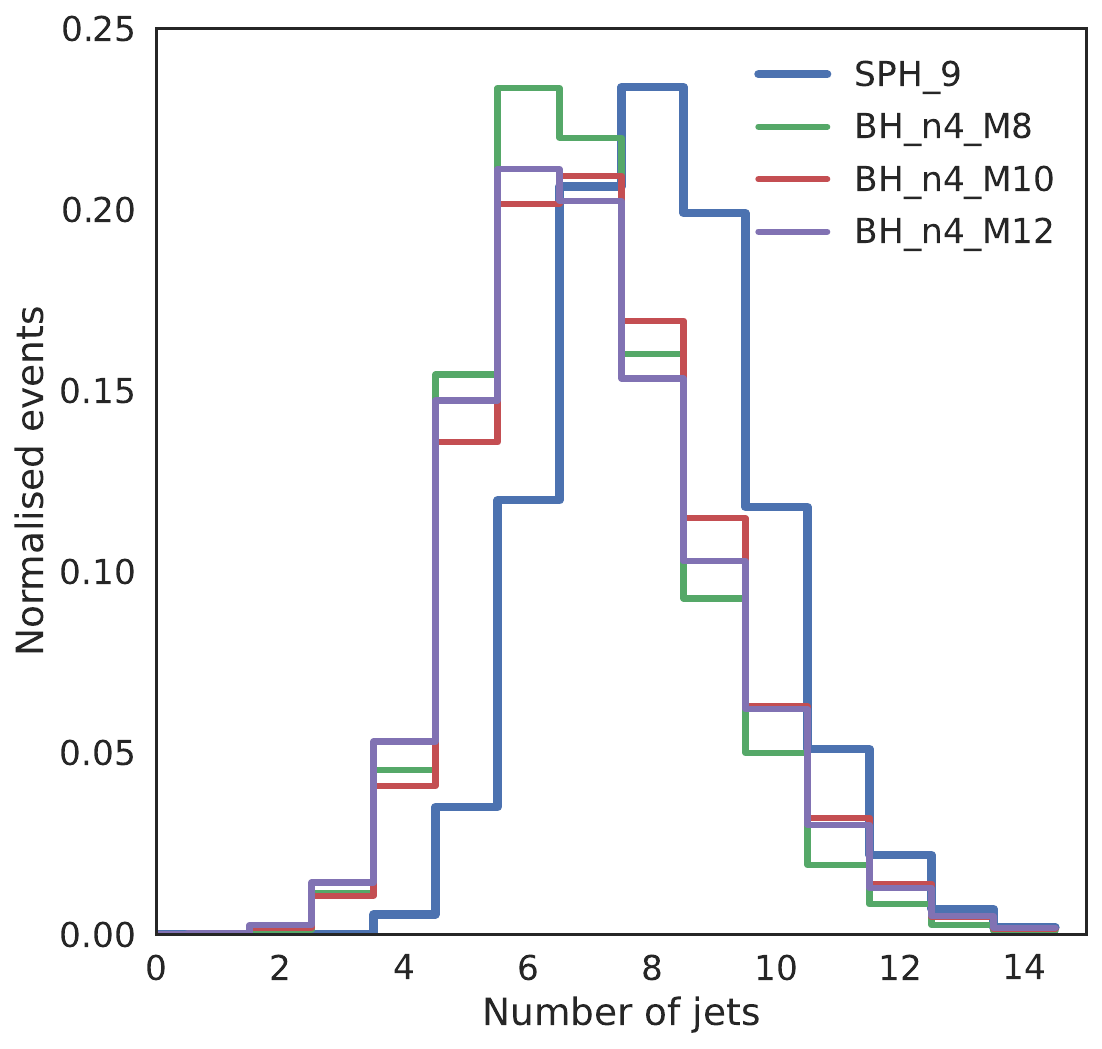}         
         \includegraphics[scale=0.25]{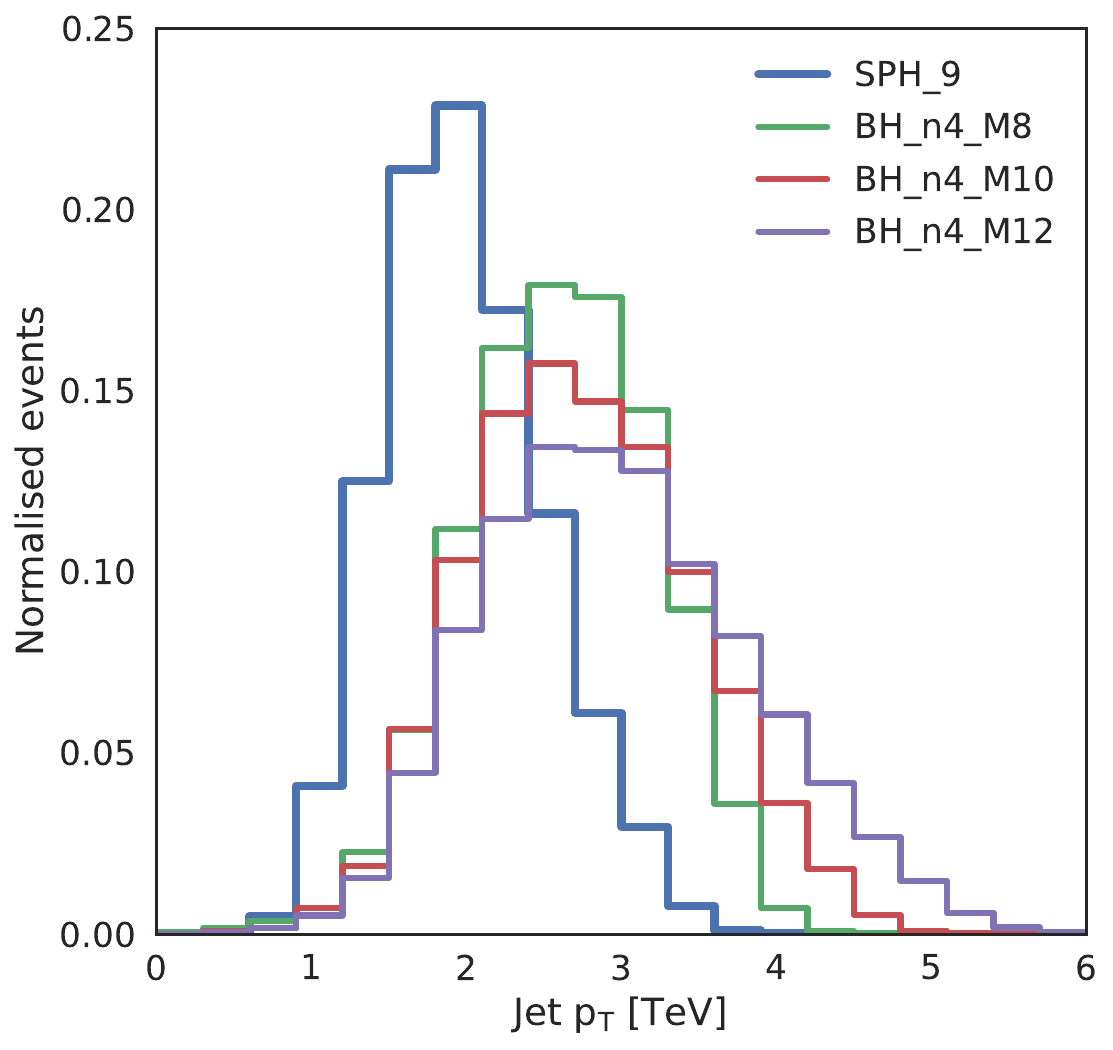}
         \includegraphics[scale=0.25]{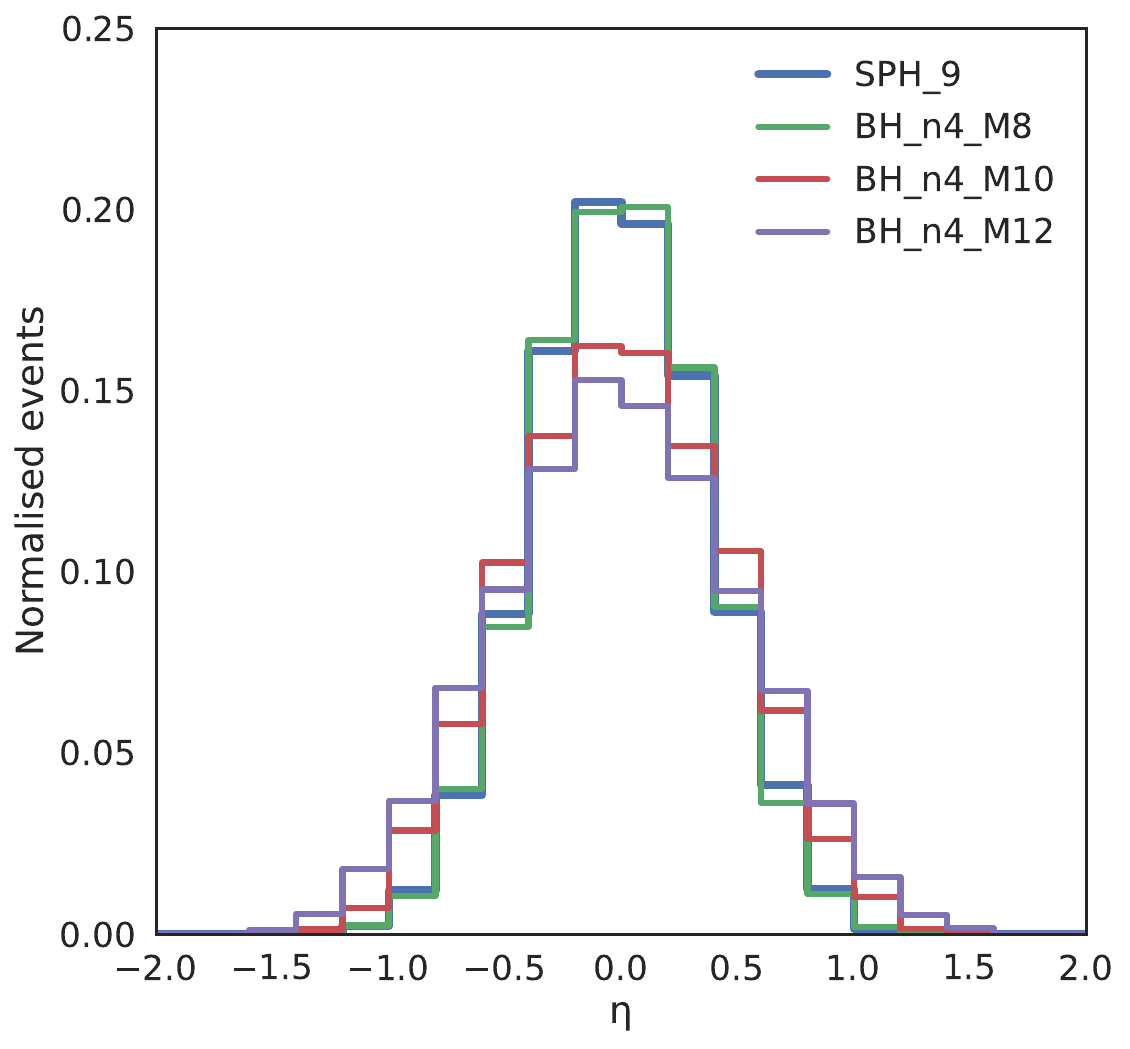}
        \caption{Normalised distributions of the number of signal jets (left),
        the $p_T$ of all signal jets (middle) and 
        the pseudorapidity $\eta$ of the highest-$p_T$ jets (right).
        The sphaleron events are shown with blue histograms.
        The plots in the upper panel show the BH scenarios with $n = 2$ (orange), 4 (red) and 6 (brown),
        while those in the lower panel show the BH scenarios with $M_{\rm min} = 8$ (green), 10 (red) and 12 (purple) TeV.}
        \label{jet}         
\end{figure}

\begin{figure}
    \centering
        \begin{subfigure}{\textwidth}
            \centering
            \includegraphics[scale=0.25]{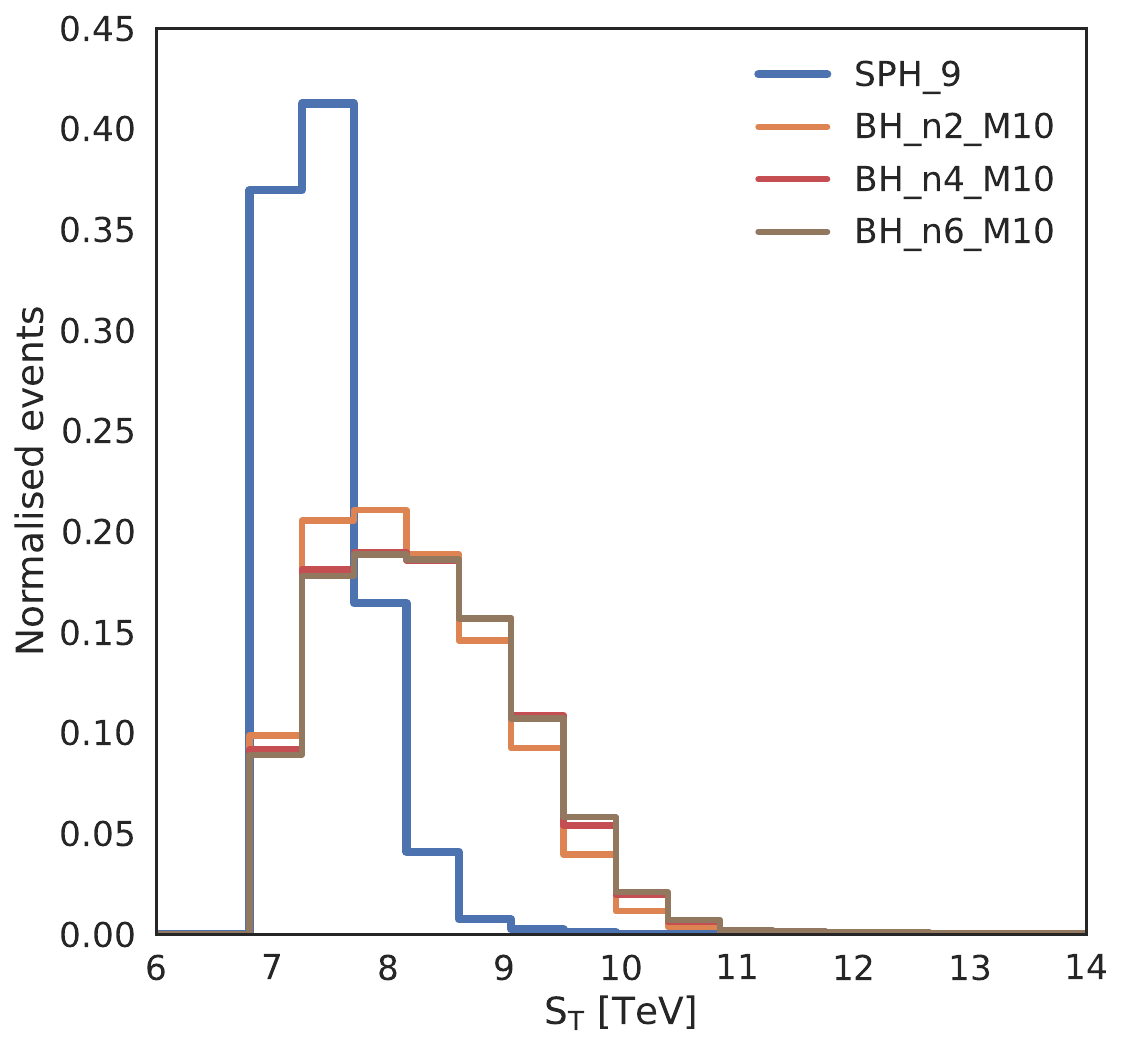}
            \includegraphics[scale=0.25]{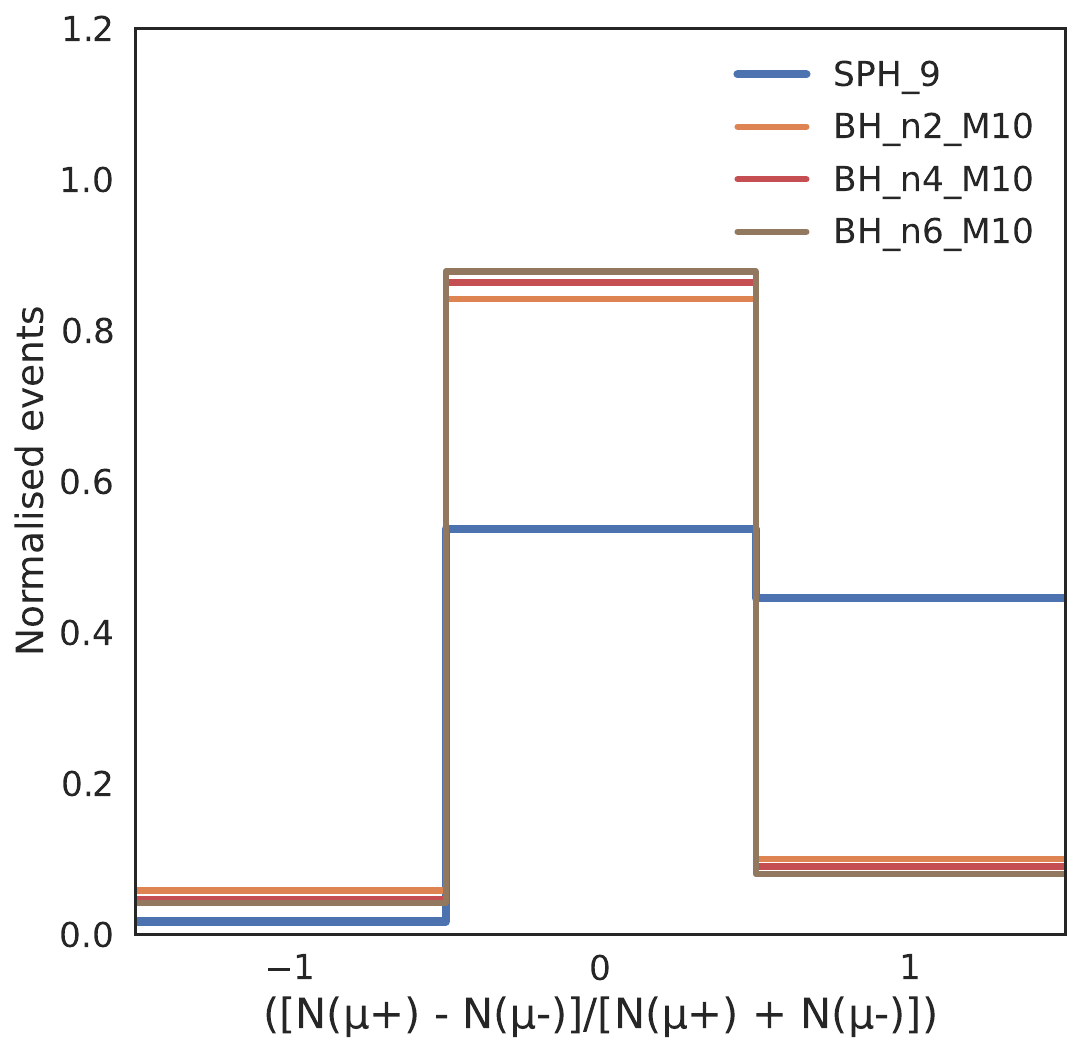}  
        \end{subfigure}
        \begin{subfigure}{\textwidth}
            \centering
            \includegraphics[scale=0.25]{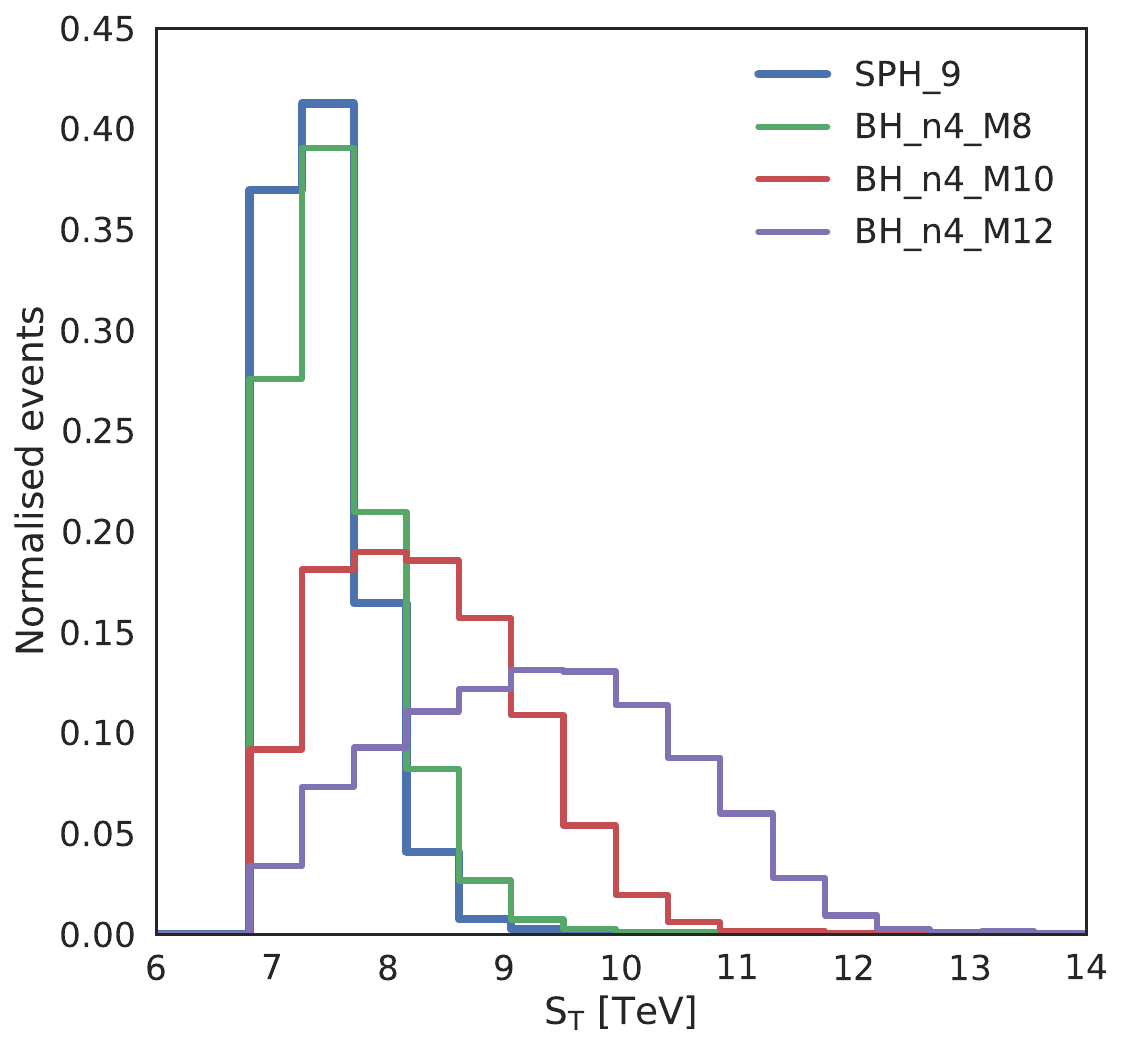}
            \includegraphics[scale=0.25]{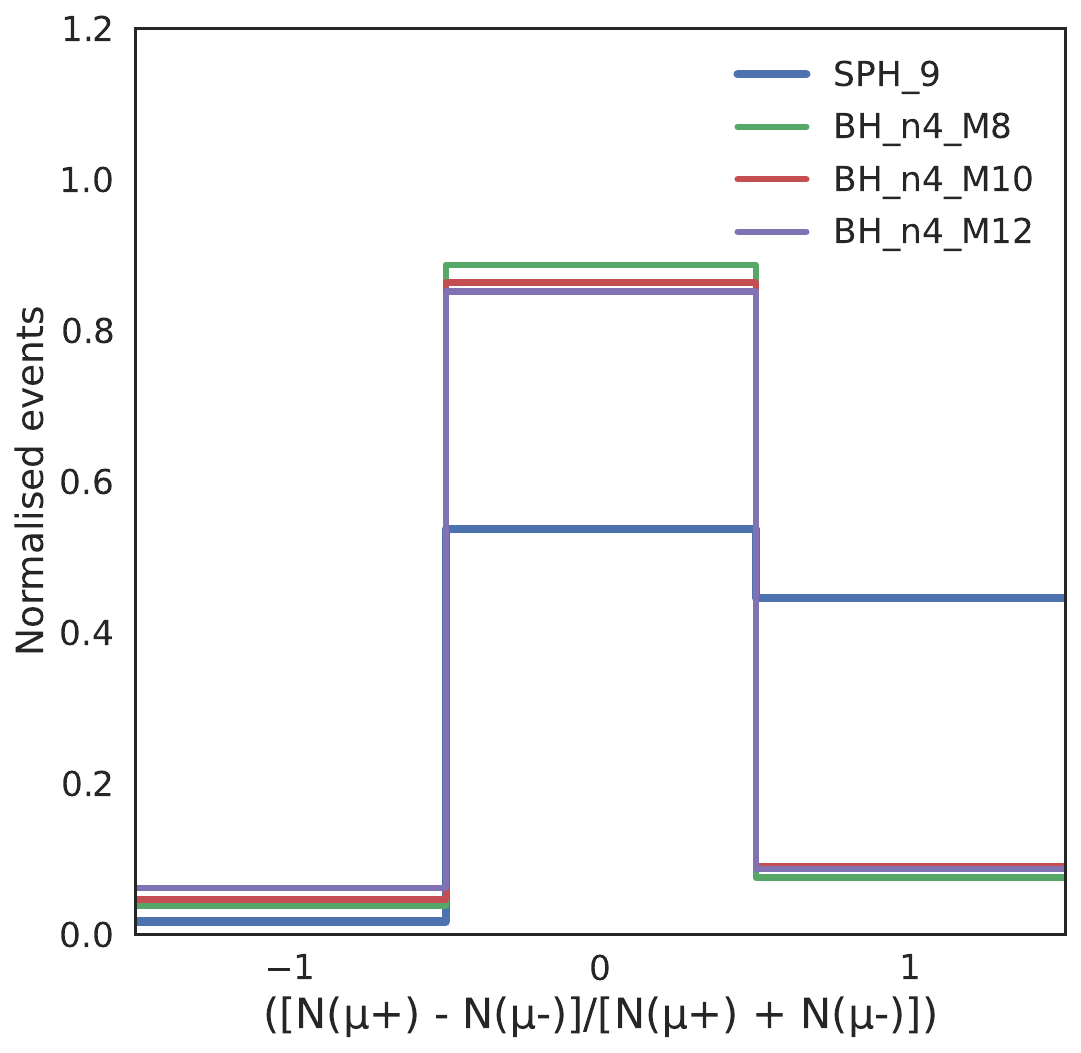}
        \end{subfigure}
    \caption{Normalised distributions of $S_T$ (left) and the muon charge asymmetry ${\rm CA}_\mu$ (right).  The same colour coding is used as in Fig.\ \ref{jet}.}
    \label{st}
\end{figure}

In order to assess the sensitivity to the model hypotheses of high-level kinematical variables, we compare their distributions in Figs.~\ref{jet} and \ref{st}.
In the upper panels of both figures
we fix the minimal BH mass at 10 TeV and vary the number of extra dimensions as $n = 2$ (orange), 4 (red) and 6 (brown),
whereas in the lower panels, we fix $n=4$ and varied $M_{\rm min} = 8$ (green), 10 (red) and 12 (purple) TeV. 
In both panels, the distributions for sphaleron events are shown in blue.

The left panel of Fig.\ \ref{jet} shows the number of signal jets, where 
we see that the sphaleron has on average 2-3 additional signal jets compared to the BH scenarios.
The majority ($\sim 90 \%$) of sphaleron events have 6-10 signal jets
and the distribution peaks at 8.
On the other hand, the distributions for BH scenarios 
peak around 5-6 jets and have a relatively long tail towards higher multiplicities.
The signal jet multiplicity is almost insensitive to the number of extra dimensions 
and the minimal BH mass.

The middle panel of Fig.\ \ref{jet} displays the accumulated $p_T$ distributions of all signal jets.
We see that the average signal jet $p_T$ is much lower for sphaleron events.
The distribution for sphaleron events has a sharper peak around 2 TeV,
whereas the distributions for BH scenarios have a broader shape, and the average jet $p_T$ is much higher, around 2.5-3.5 TeV.
We observe in the upper middle plot that the jet $p_T$ for $n=2$ BH is slightly lower than that for $n=4$ and 6, on average.
In the lower middle plot, we also see that the jet $p_T$ distributions have longer tails in the higher-$p_T$ region for larger minimal BH masses. 

The distributions of the pseudorapidity $\eta$ of the highest-$p_T$ jet are shown in the right panel of Fig.\ \ref{jet}.
As can be seen, the most energetic jet is produced in the central region, $|\eta| \lesssim 1$, both for sphaleron and BH events.
The $\eta$ distributions are similar among all scenarios,
though one can observe that the sphaleron and the $(M_{\rm min}, n) = (8\,{\rm TeV}, 4)$ BH have slightly narrower distributions, compared to the other BH scenarios.  

The left panel of Fig.\ \ref{st} shows the $S_T$ distributions.
As can be seen, the $S_T$ distribution for sphalerons sharply peaks around 7.5\,TeV. 
We observe that the $S_T$ distribution for BH scenarios is sensitive to $M_{\rm min}$, but not to the number of extra dimensions, $n$.
We see in the lower-left plot that the $S_T$ distribution for the $(M_{\rm min}, n) = (8\,{\rm TeV}, 4)$ BH is similar to that for sphalerons.
The distribution gets broader and shifts towards higher energies as the minimal BH mass increases.  
The $S_T$ distribution peaks around 9.5 TeV for the $(M_{\rm min}, n) = (12\,{\rm TeV}, 4)$ BH scenario.

The right panel of Fig.\ \ref{st} displays 
the charge asymmetry of signal muons, defined by the difference between the numbers of $\mu^+$ and $\mu^-$ in an event, divided by the total number of muons:
${\rm CA}_\mu \equiv (N_{\mu^+} - N_{\mu^-})/(N_{\mu^+} + N_{\mu^-})$.
If the event contains no signal muons, we set ${\rm CA}_\mu = 0$.
We see in the plots that there is a large excess of $\mu^+$ in sphaleron events:
sphaleron events have ${\rm CA}_\mu = 0$ or 1 with $\sim 50$ \% chance.
This asymmetry is expected because in the quark-quark initial state,
the sphaleron vertex involves exactly one anti-muon or 
exactly one anti-muon-neutrino.
Extra \mbox{(anti-)muons} may arise from decays of anti-top-quark or heavy hadrons.
Contrary to the sphaleron case,
the majority ($\sim 90$ \%) of BH events have zero muon charge asymmetry.
This is expected because semi-classical BH decays are essentially a thermal process.
In the plots, we however observe a small excess for $\mu^+$ in
all BH samples.
This is a consequence of the charge conservation of BH decays 
and the proton contains more up-quarks than down-quarks, especially in a large-$x$ region of the parton distribution function. 
Although the muon charge asymmetry has different distributions between BH and EW sphaleron events, it is not the best discrimination variable, particularly when the available data size is small, as the majority of events have ${\rm CA}_\mu = 0$.

Observed differences in several kinematical distributions 
between sphaleron and various BH scenarios suggest that the separation 
between sphaleron and BH scenarios should be relatively straightforward 
with sufficient collider data, though the optimal discrimination method is not clear. 
Separation between different $M_{\rm min}$ among BH scenarios 
seems also possible with some resolution, since 
the jet $p_T$ and $S_T$ distributions depend, to some extent, on $M_{\rm min}$.
On the other hand, 
separation of the number of extra dimensions $n$ 
seems challenging as 
no distribution in Figs.\ \ref{jet} and \ref{st} exhibits a clear dependence on $n$.
In the following sections, we investigate how well one can distinguish different scenarios, 
with a given number of observed events,
using modern ML techniques. 

\section{Machine Learning setup}
\label{sec:ML}

In this study, we examine three types of ML methods. The first two are based on the well-established XGBoost algorithm \cite{Chen2016}, which creates an ensemble of decision trees to effectively separate different events. The difference between the two models based on XGBoost is the choice of input features. In the first approach, we use high-level variables based on reconstructed objects, while in the second we utilise low-level tracker and calorimeter signals. The third method uses the CNN-based ResNet \cite{He2015} architecture, which is known to be very powerful in image recognition applications.

The ML models are all trained and evaluated using shared input data consisting of 10000 (training), 3000 (validation) and 15000 (testing) events for each 
physics model hypothesis, giving a total of 60000, 18000 and 90000 events, respectively. These are the number of events after cuts are applied, so the models are trained on a balanced dataset. The models 
are trained using the training set, fine-tuned using the validation set and finally, the networks are 
used to predict the labels of the independent test set events.
The predicted and true labels of the test set are then used to calculate the 
metrics for evaluation. 
For this study, the metric chosen is a simple global accuracy (ACC), i.e., the number of correctly labelled events divided by the total number of events, 
such that perfect labelling of all events corresponds to $\rm{ACC}=1.0$, and a random classifier 
results in $\rm{ACC}=1/n$, where $n$ is the number of distinct classes (under the assumption that the dataset is balanced, i.e., all classes are equally represented).
\footnote{Code, results and data specifications are openly available on our project GitHub \texttt{https://github.com/choisant/imcalML} \cite{aurora_2023_10033266}} 

To assess the uncertainties due to the randomness of the ML training, 
for each of the three ML methods, we create five independent classifiers (decision trees for XGBoost and neural networks for ResNet) by shuffling the training data at the start of each training.
The mean global accuracy and its standard deviation are estimated with these five independent classifiers.  

\subsection{XGBoost}
\label{sec:XGBoost}

The XGBoost algorithm was developed during the highly successful Higgs ML 2014 Kaggle competition (\hyperlink{https://www.kaggle.com/c/higgs-boson}{www.kaggle.com/c/higgs-boson}) \cite{Adam-Bourdarios_2015}. It is a regularizing gradient boosting framework library and enables fast training of decision tree models. XGBoost is implemented in many modern HEP analyses such as the diphoton search from ATLAS \cite{ATLAS:2021jbf}. As such, XGBoost is a state-of-the-art machine-learning tool for high-energy particle physics and is expected to yield good results in a short training time.

In this study, XGBoost models with two different input types have been explored.
The first XGBoost method (XGBoost-High) uses the high-level inputs 
of the reconstructed objects. 
The selected input features are\footnote{The jets and leptons are ordered by descending} $p_T$. the first eight jets ($p_T^{\rm miss}$, $\eta$, $\phi$), the first two electrons and muons ($p_T$, $\eta$, $\phi$) and the missing transverse energy ($p_T^{\rm miss}$, $\phi$), which comprises 38 input features.
The XGBoost algorithm tolerates missing values, but the data must be the same length for all events.  
The data are padded with \texttt{NaN} if an event does not have enough jets or leptons.

In the second XGBoost method (XGBoost-Low) low-level input is used.
In this case, the input features consist of the signals in the tracking system, the electromagnetic calorimeter (ECAL) and the hadronic calorimeter (HCAL). The calorimeter hits, also referred to as calorimeter towers, represent the summed-up energies at discrete angular positions across the detector. This means that the radial coordinates where the energy was deposited are not retained. 
The 30 highest-energy hits in each of the three different detector systems, and their corresponding angular position, were selected. This sums to 270 input features in total. Although we experimented with other variations of input features, 
we did not find any improvement by adding more features with lower-energy hits.

\subsection{Convolutional neural network}
\label{sec:CNN}

CNNs are neural networks with convolutional layers. 
It is known that this architecture works extremely well in recognizing patterns in image-type data, that is data which have a matrix structure where the relative positioning of each input in the matrix is important.
In this study, we use the ResNet18 architecture with 18 layers\footnote{We tested models with more layers, but did not find an increase in performance.}, implemented with the PyTorch package\cite{paszke2019pytorch}. We convert low-level collider event data into event images as follows:
starting from the 2D angular coordinates of the calorimeter towers and tracks, $(\phi, \eta)$,
we divide the range $\phi \in [-\pi, \pi], \eta \in [-5, 5]$ into $50 \times 50$ equal sized bins.\footnote{Higher resolutions than $50 \times 50$ were tested. We found either no improvement or slightly worse performance.} The total energy deposit (in GeV, not normalised) in the ECAL, HCAL and tracks in a given bin is assigned to their respective layer in that bin. The result is a 3-layer 2D histogram with the input shape $50\times50\times3$, which we refer to as an event image and has 7500 input features in total. 
In the appendix, we show example event images of each of the physics scenarios in Fig.\ \ref{event_image}.

In the course of the training process, data augmentation is applied to the images in order to artificially increase the statistics of the dataset. The augmentation consists of random rotations in the $\phi$-direction and random flipping of the $\eta$ axis. These transformations do not affect the physics of the events, because of the symmetries of the detector and the collision setup. The images in the validation and testing set are not transformed in any way. 

During the training, random batches of 128 images are loaded into memory at a time. Data augmentation is applied to each image, and then the network is trained using an exponential cyclic learning rate scheduler \cite{smith2017} with a base learning rate of 0.001, a maximum learning rate of 0.01, and gamma = 0.85 and the Adam optimization algorithm \cite{Adams2014} based on the calculated cross-entropy loss from the batch. The network is trained for 40 epochs.

\section{The classification performance}
\label{sec:results}

\begin{table}
\begin{center}
\begin{tabular}{ c c c c } 
 \hline
 \hline
Model & XGBoost-High & XGBoost-Low & ResNet18 \\
 \hline
 Accuracy & $48.91 \pm 0.05 $ & $46.69 \pm 0.08$ & $\boldsymbol{57.92 \pm 0.37}$  \\ 
 Training time &\textbf{2 min} & 3.5 min & 8.5 min  \\ 
 Testing time & 10 s & \textbf{7 s} & 31 s \\ 
 \hline 
 \end{tabular}
\caption{The global accuracy (in percent) as well as the training and testing time for XGBoost-High, XGBoost-Low and ResNet18 ML methods. The mean and error of the global accuracy were estimated from an ensemble of five separately trained classifiers for each ML method.
The quoted training and testing times were measured based on a GPU with an NVIDIA RTX A4500 graphic card.
}
 \label{resultstable}
\end{center}
\end{table}

The mean global accuracy with uncertainties due to the stochastic ML training process, as well as training and testing times for each ML method,
is presented in Table \ref{resultstable}. 
The three ML methods we examined have global accuracy classifier scores ranging from $46.69 \pm 0.08$ (XGBoost-Low) to $57.92 \pm 0.37$ (ResNet18).
It is interesting that ResNet18, which uses only low-level data, can perform significantly better than XGBoost does with high-level data inputs. It should be noted that the training and testing times for the CNN model are approximately 3-4 times longer than for XGBoost. This is a natural effect of the deep neural network's complex structure and the large difference in the number of input features.

\begin{figure}
     \centering
     \includegraphics[width=0.32\textwidth]{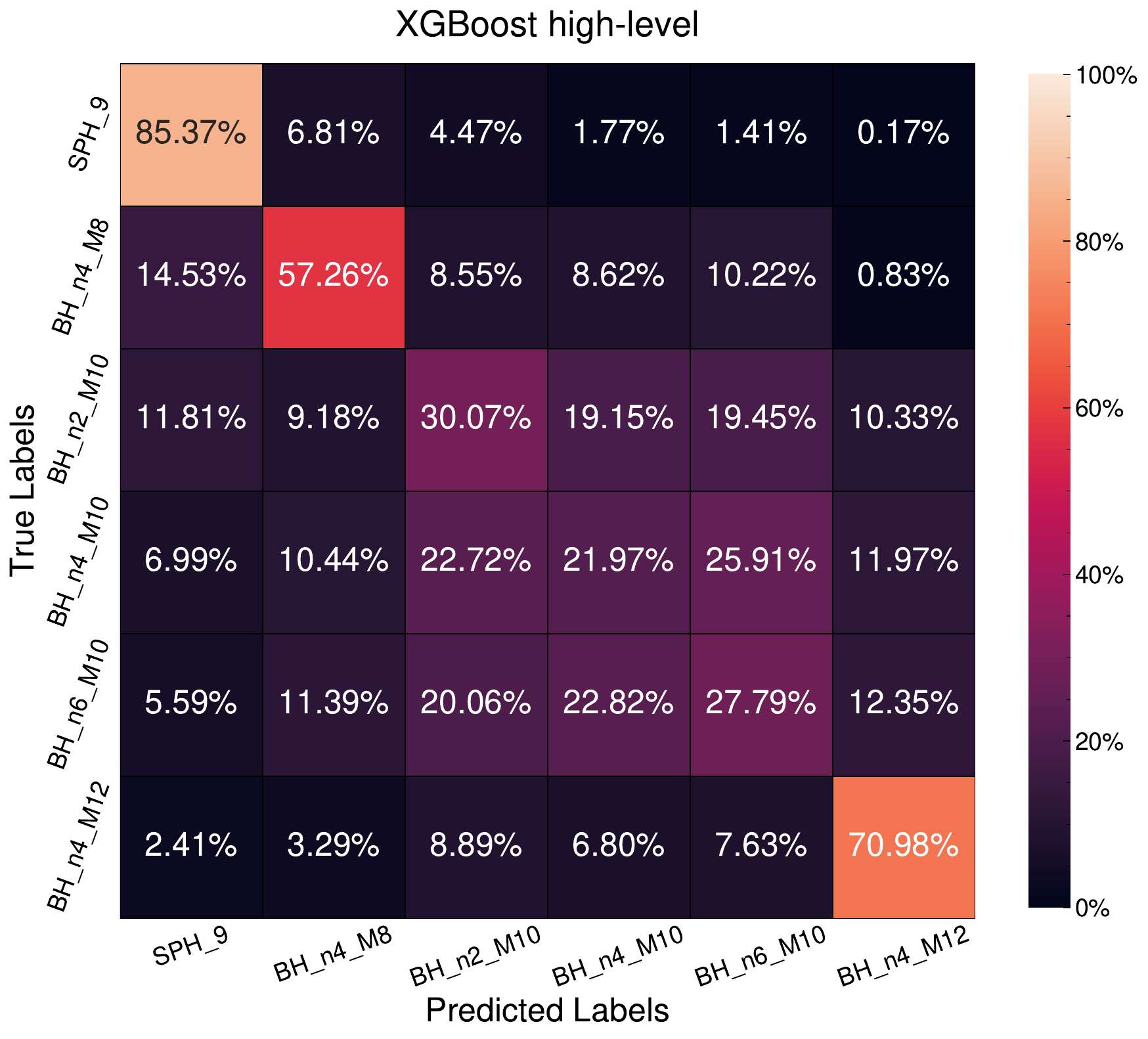}
     \includegraphics[width=0.32\textwidth]{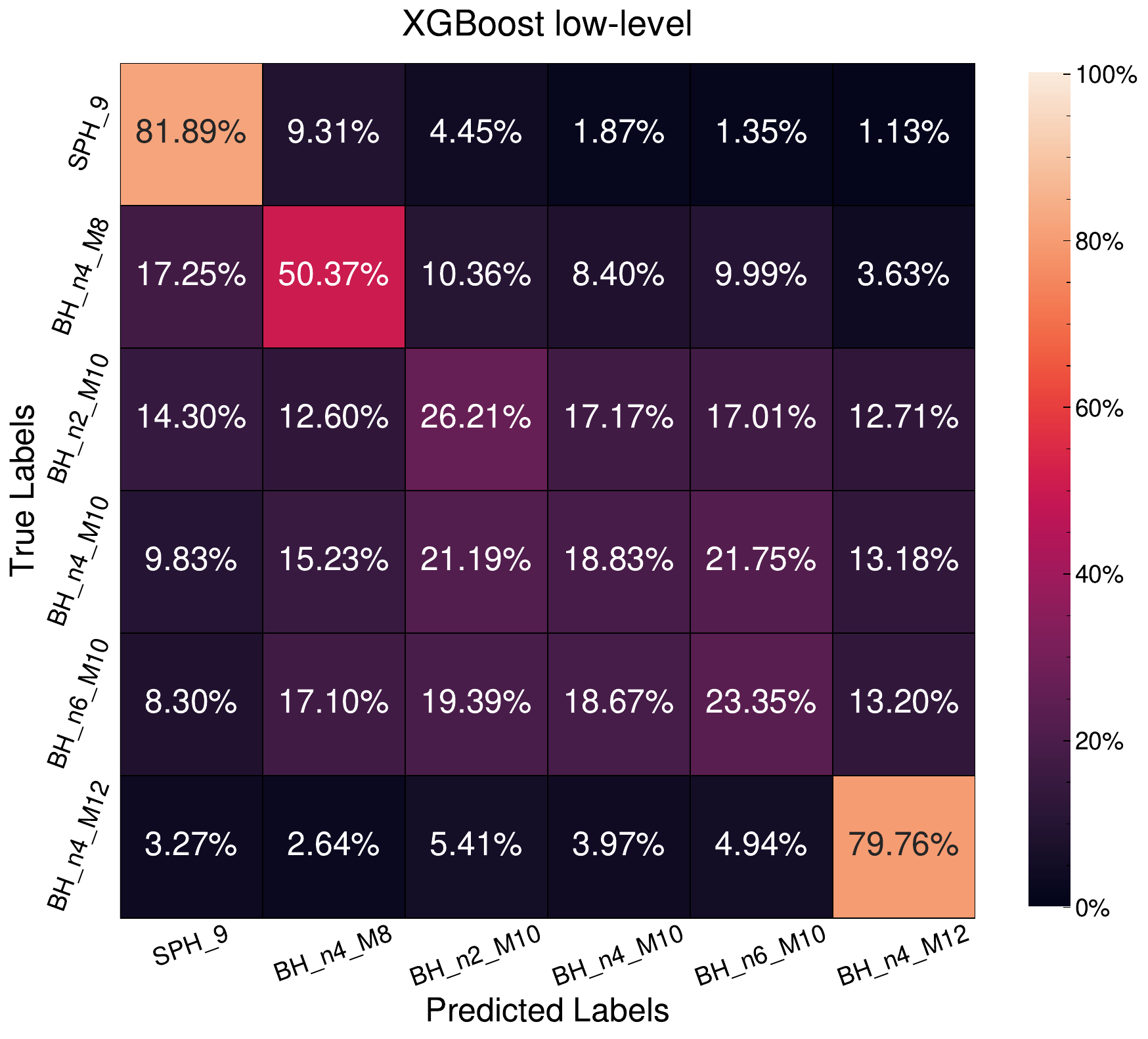}
     \includegraphics[width=0.32\textwidth]{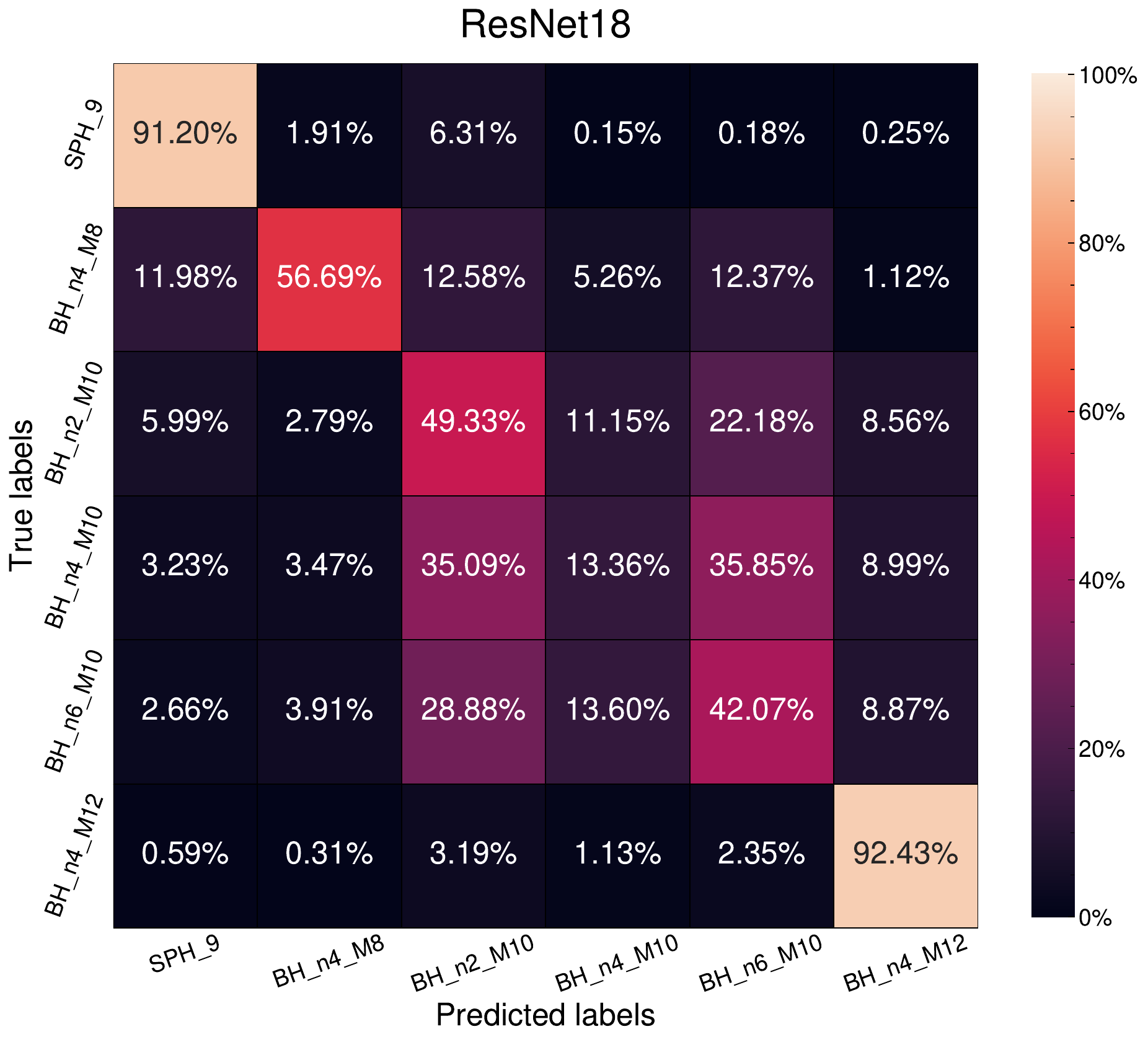}
     \caption{Confusion matrices for the XGBoost-High (left),
         XGBoost-Low (middle) and CNN ResNet18 (right). }
     \label{conf_cnn}
\end{figure}

Fig.\ \ref{conf_cnn} shows the confusion matrices of the three ML methods we examined. The separation of sphalerons and BHs is the objective that all models handle well. In particular, ResNet can correctly identify the sphaleron events with $\sim 91$\% accuracy.
Furthermore, when separating the different BH samples, the models are much better at distinguishing samples with different minimal BH masses, than with different numbers of extra dimensions. ResNet correctly identifies 92\% of all the black holes with a minimum mass of $12\,$TeV, compared to 71\% and 80\% respectively for XGBoost-High and XGBoost-Low. ResNet is also fairly good at separating the different number of dimensions. It correctly classifies 49\% of the events with 2 extra dimensions and 42\% of the ones with 6 extra dimensions. 

To demonstrate the practical application of the ML classifiers, we perform hypothesis tests using Poisson statistics, for a given number of observed signal events, $N_{\rm obs}$, in our signal region ($N \ge 5$, $S_T \ge 7$\,TeV). We have conducted this exercise only with the ResNet model as it outperforms XGBoost. 

For each observed event $i$ of model hypothesis scenario $J$,\footnote{There is no case in which the real LHC data could contain a mix of scenarios.} ResNet assigns a classification label, $L_i$, where $J$ and $L_i$ belong to the set of six model hypotheses in Table \ref{hypotheses}.
Using the test set of 15000 events, we first create a normalised template distribution of classification labels $L$ for each scenario $J$. Roughly speaking, each bin of the distribution, $P_J(L)$, can be interpreted as the probability that an observed event of scenario $J$ is classified as scenario $L$ by the ResNet.

Iterating over all the possible scenarios where the data corresponds to hypothesis $I$, we test each hypothesis $J$ to see if it can be excluded. We define the label $L_J^*$ into which the event of $J$ is most likely to be classified by the ResNet: i.e.,\ $P_J( L_J^* ) \geq P_J( L )$ for all $L$. 
If $N_{\rm obs}$ events are observed in the signal region, the average number of events labelled as $L_J^*$ is $\lambda = N_{\rm obs} \cdot P_J( L_J^* )$.  
If hypothesis $J$ is correct, the probability of
observing $n$ events labelled as $L_J^*$ is given by
the Poisson function  
\be
P( n | \lambda ) = \frac{ \lambda^n e^{-\lambda} }{n!} \,.
\ee
The hypothesis $J$ will be excluded at $95\%$ confidence level if the number of observed events labelled as $L_J^*$, $N(L_J^*)$, in the data is too few: i.e.,
\be
p_J \equiv \sum_{n=0}^{N(L_J^*)} P(n|\lambda) < 0.05 \,.
\ee

\begin{figure}[t!]
     \centering
     \includegraphics[width=0.32\textwidth]{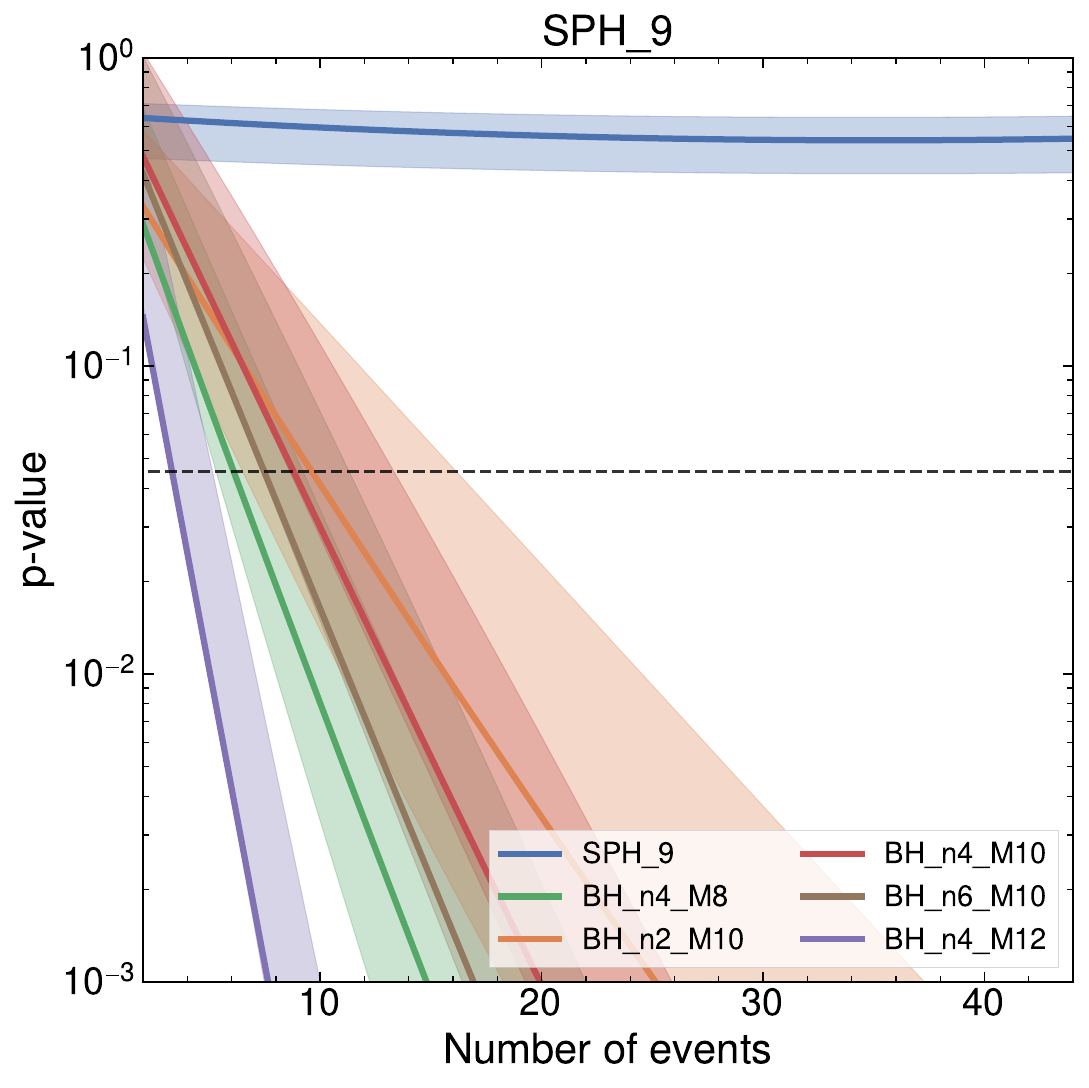}
     \includegraphics[width=0.32\textwidth]{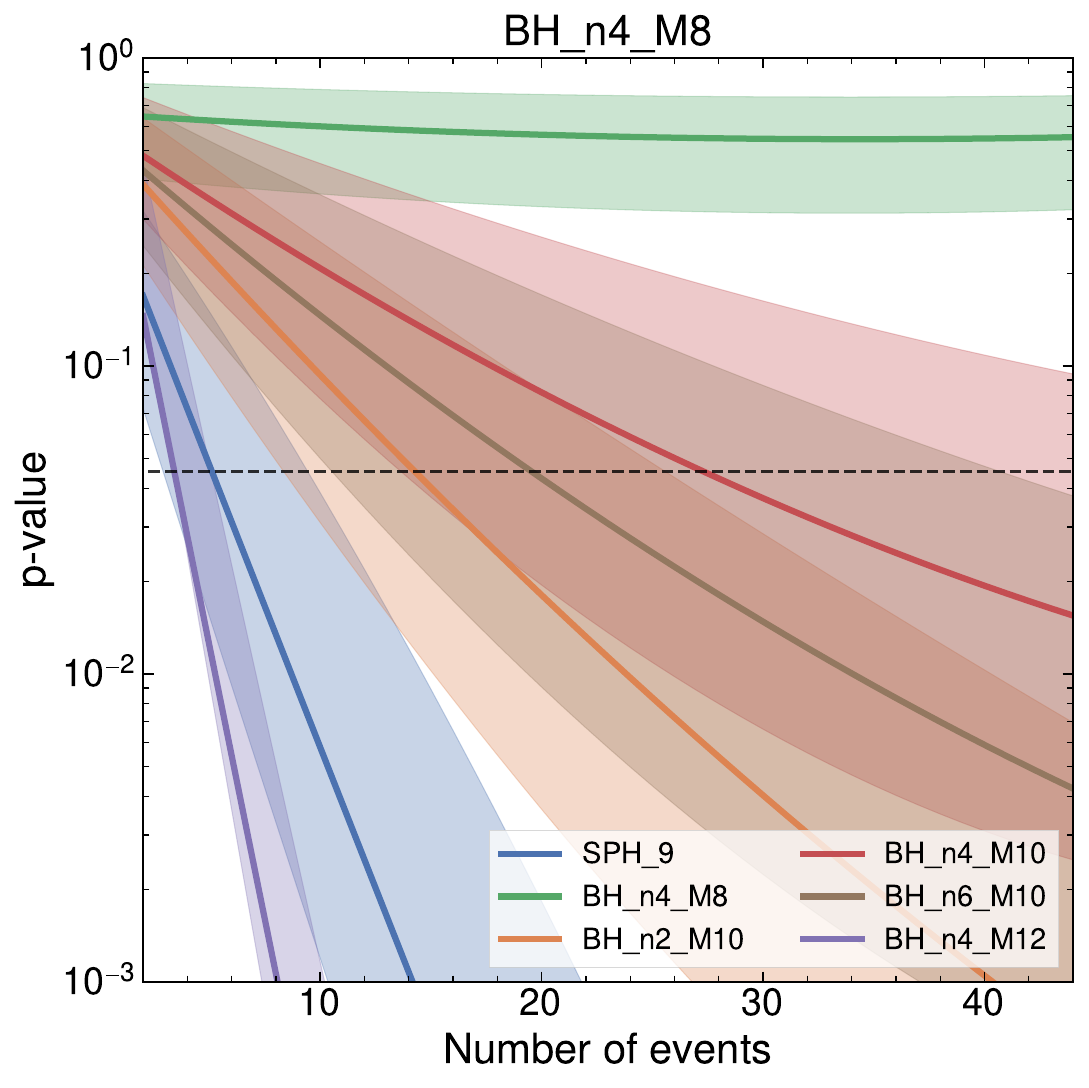}
     \includegraphics[width=0.32\textwidth]{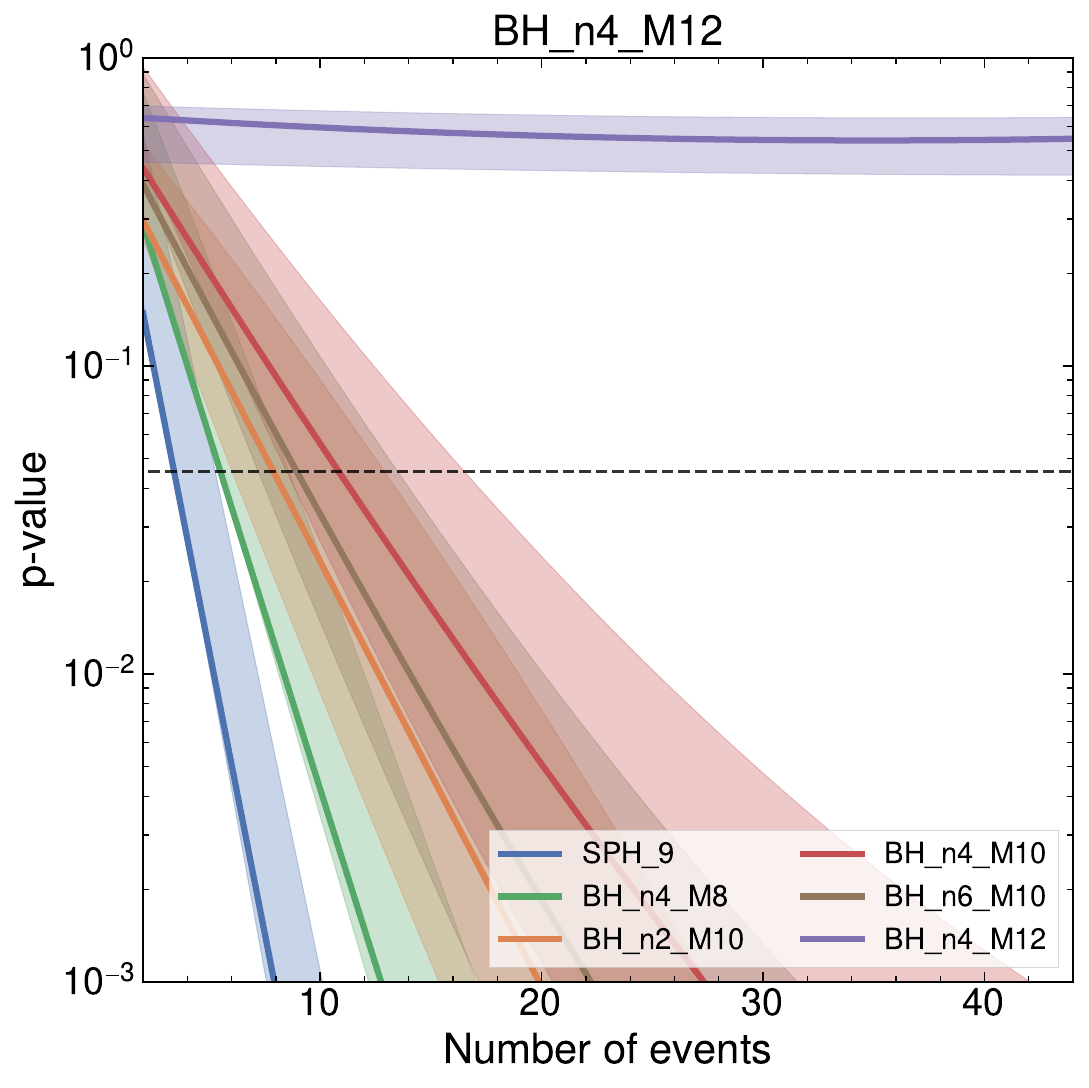}
     \includegraphics[width=0.32\textwidth]{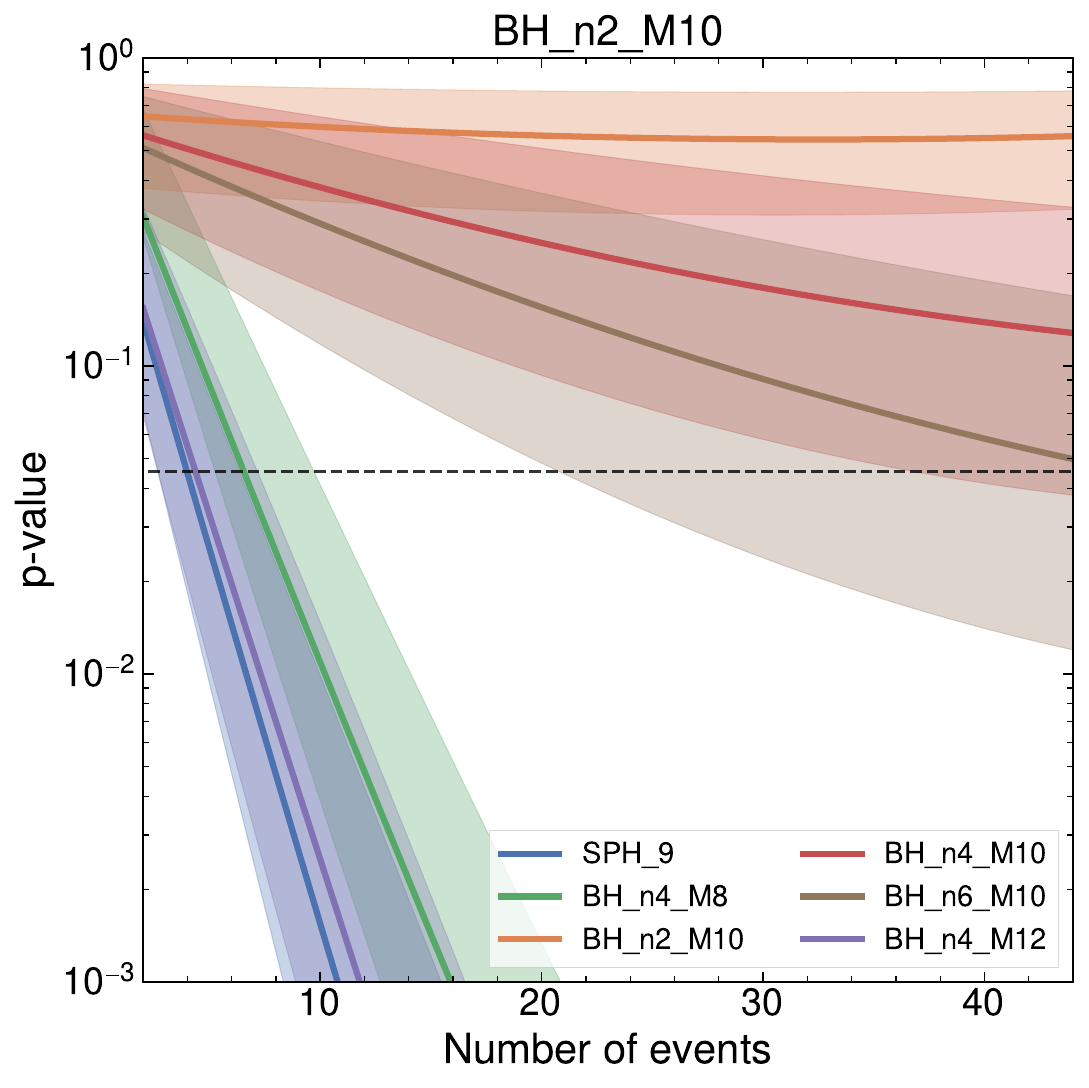}
     \includegraphics[width=0.32\textwidth]{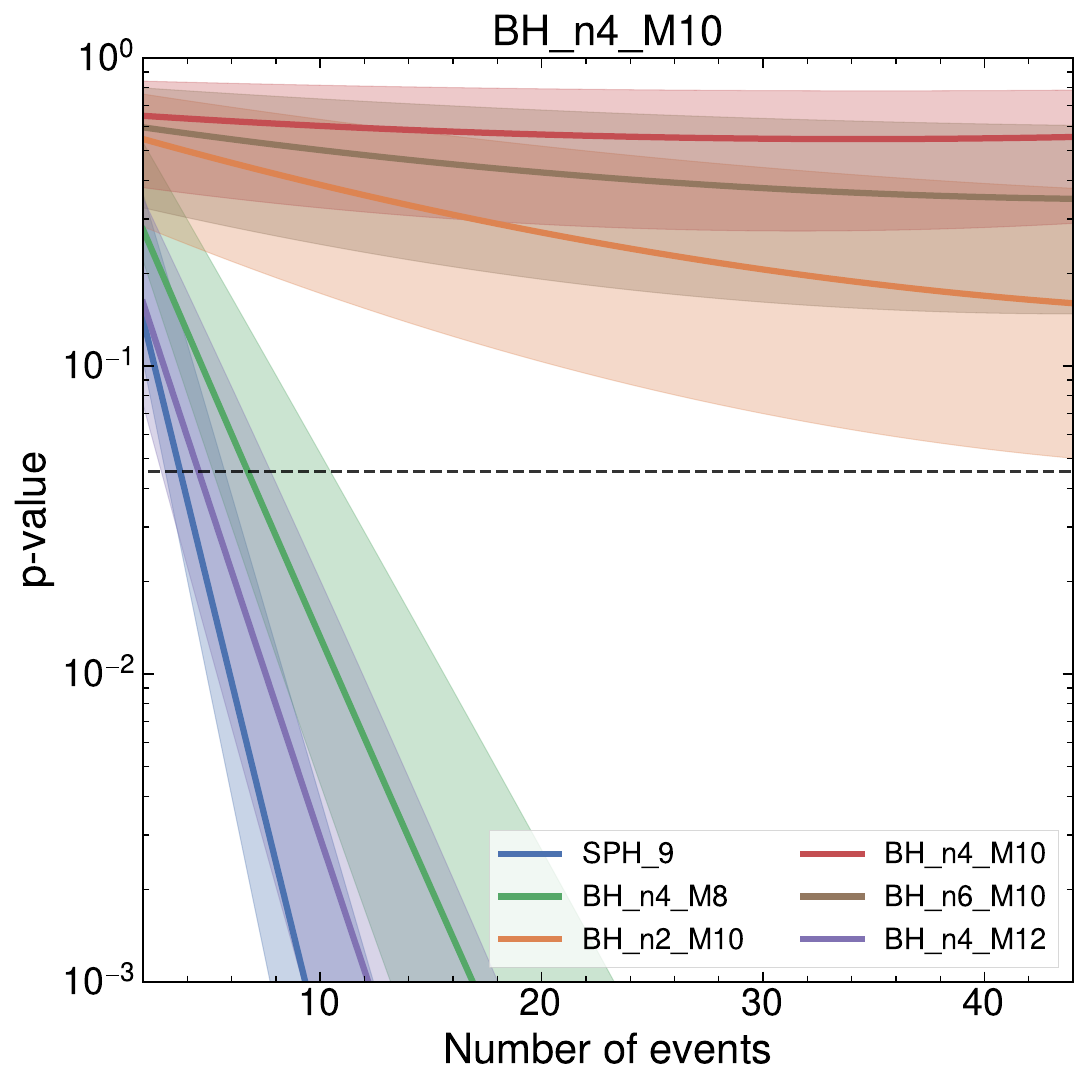}
     \includegraphics[width=0.32\textwidth]{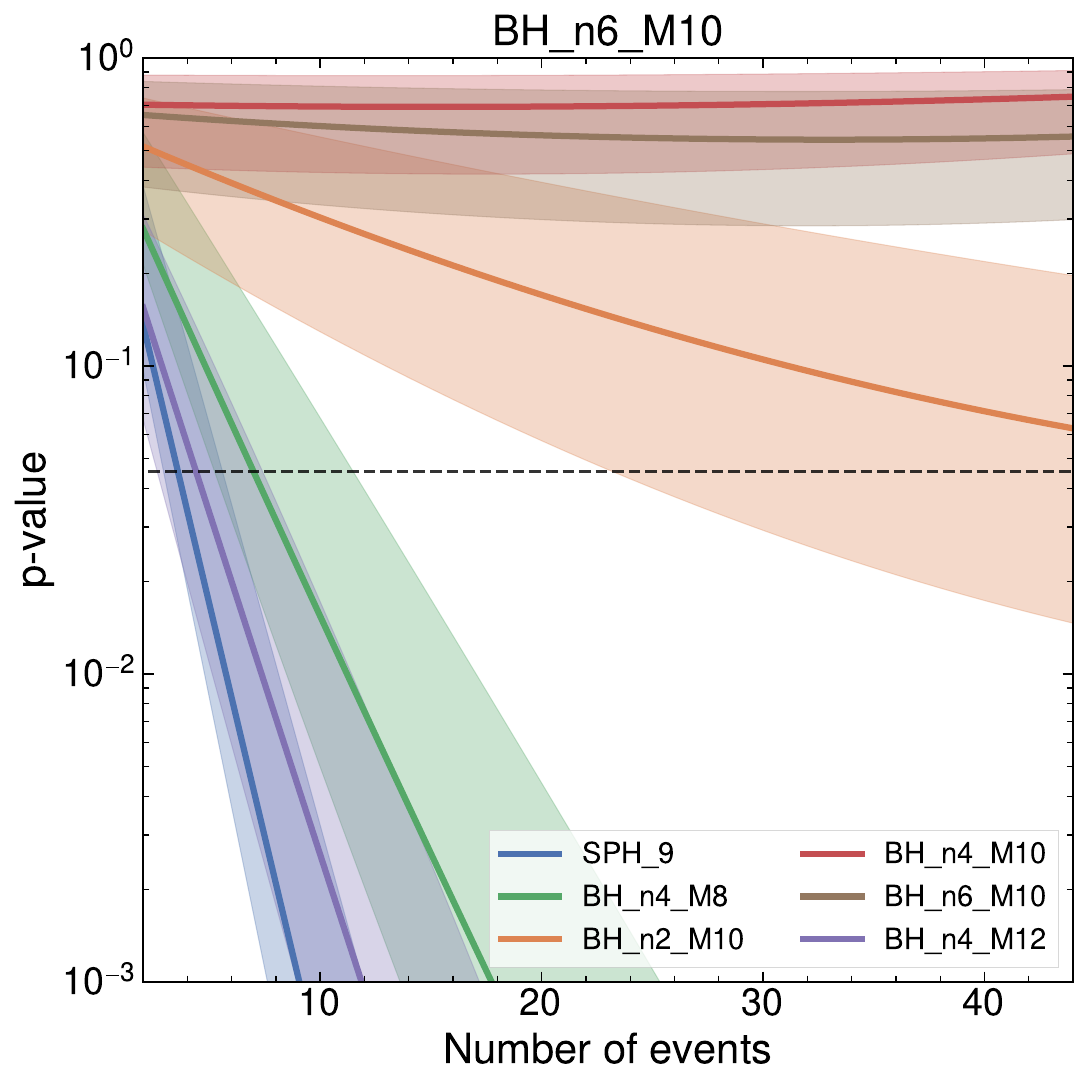}
     \caption{The exclusion $p$-values of six model hypotheses calculated using Poisson statistics:
     SPH\_9 (blue), BH\_n4\_M8 (green), BH\_n2\_M10 (orange), BH\_n4\_M10 (red), BH\_n6\_M10 (brown), BH\_n4\_M12 (purple) as a function of the signal events registered in the signal region.
     The true hypotheses are indicated at the top of each plot.
     The horizontal black-dashed line represents the 2-$\sigma$ exclusion.}
     \label{p_values}
\end{figure}

Given the current limits on the BH and EW sphaleron cross sections, \cite{CMS:2018ozv, Ringwald:2018gpv},
$\sigma \lesssim 0.1$\,fb, we do not expect that a large number of BH/sphalerons events could be collected in LHC Run-2, -3 and HL-LHC. For a small number of signal events, the corresponding $p$-values can fluctuate significantly.
To understand the magnitude of the fluctuation, we performed 3000 pseudo-experiments, sampled from the 15000 event test set, for each $N_{\rm obs}$ of the  ``true'' scenario and estimated the upper and lower errors on the $p$-values.   

In Fig.\ \ref{p_values}, we show the result of our $p$-value estimation as a function of the number of observed events in the signal region.  
In the top-left plot, the true scenario (the origin of signal events) is SPH\_9 (EW sphaleron scenario with 9 TeV threshold energy).
The blue curve and the band around it show 
the $p$-value for the SPH\_9 hypothesis and the standard deviation, estimated from the pseudo-experiments, respectively.  
As expected, the $p$-value is ${\cal O}(1)$ and does not decrease as the number of signal events increases, since it is the correct hypothesis.    
The other curves show the $p$-values for various BH hypotheses:
BH\_n4\_M8 (green), BH\_n2\_M10 (orange), BH\_n4\_M10 (red), BH\_n6\_M10 (brown), BH\_n4\_M12 (purple).
The horizontal dashed black line in the plot corresponds to the 2-$\sigma$ ($p \simeq 0.05$) exclusion.
Observing less than 15 events is often enough to exclude all of our BH hypotheses at more than the 2-$\sigma$ level. 

The top-middle and top-right plots show the $p$-values of various hypotheses when the true scenarios are BH\_n4\_M8 and BH\_n4\_M12, respectively.
In both cases, we can observe that the $p$-values for the correct hypotheses remain ${\cal O}(1)$ for a larger number of signal events.  
We also notice that the sphaleron hypothesis and the BH hypothesis with $M_{\rm min}$ differ by 4 TeV from the correct hypothesis (i.e.,\ BH\_n4\_M12 (purple) in the middle plot and BH\_n4\_M8 (green) in the right plot), are excluded very quickly while accumulating signal events.  
For those scenarios, a small number of events are sufficient for a 2-$\sigma$ exclusion.
In the same plots, we see the $p$-values for the BH hypotheses whose $M_{\rm min}$ differs by 2 TeV from the correct hypothesis. 
These are the scenarios with 
$M_{\rm min} = 10$ TeV and $n=2$ (orange), 4 (red) and 6 (brown).
We see that the $p$-values of these three hypotheses decrease at a somewhat slower rate.

The three plots at the bottom of Fig.\ \ref{p_values}
show the $p$-values when the true scenarios have $M_{\rm min} = 10$\,TeV
and $n=2$ (bottom-left), 4 (bottom-middle) and 6 (bottom-right).
The sphaleron hypothesis can be excluded with 
a small number of events at the 2-$\sigma$ levels.
The exclusion of the BH scenario with $M_{\rm min} = 12$\,TeV (purple) is easier than that with $M_{\rm min} = 8$\,TeV (green).
As can be seen from these three plots, the discrimination between the same $M_{\rm min}$ and different $n$ is more challenging. 2-$\sigma$ discrimination between $n=2$ and $n=6$ requires $\sim 45$ events. 

\section{Conclusions}
\label{sec:concl}

Large-multiplicity final states of jets and leptons at the LHC are ubiquitous in exotic processes, both within the Standard Model and within its extensions.  
In particular, final states with ${\cal O}(10)$ jets plus a few leptons are expected in thermal decays of semi-classical Black Holes (BHs), which are anticipated to be produced in models with large extra dimensions, characterised by two main parameters:
the number of extra dimensions and the minimal mass of the BH.
Similar final states are also expected in the EW sphaleron/instanton-induced processes.
In this study, we have investigated whether one can discriminate among
the sphaleron scenario and five different BH scenarios listed in Table~\ref{hypotheses},
through events collected at the LHC, using modern Machine Learning (ML) methods. 

In studying kinematical distributions we observed 
that sphalerons have significantly different distributions than
those of various BH scenarios, in the number of jets, $p_T$ of jets, as well as the muon charge asymmetry.
The $S_T$ and jet $p_T$ distributions also exhibit a dependence on the minimal BH masses.
Despite these differences, 
the optimal model separation method, which works for low signal statistics, is unclear due to the complexity of large-multiplicity final states. 
On the other hand, no clear sensitivity to the number of extra dimensions was found in any of the distributions shown in Figs.\ \ref{jet} and \ref{st}. 

In this study, three ML models have been examined:
XGBoost with low- and high-level inputs (XGBoost-Low and XGBoost-High) and a convolutional neural network ResNet model.
In the latter, the input was the low-level detector information, converted into three-layer binned event images, corresponding to the signals in the ECAL, HCAL and tracking system. 
We found that the discrimination power is the highest for the ResNet model as it outperforms the state-of-the-art XGBoost method by achieving a 9\% higher global accuracy score on our test set. This is quite remarkable and highlights the capacity of neural networks to utilize low-level, high-dimensional data.

To assess the practical applicability of the best-performing ML classification, we evaluated the exclusion $p$-value of each hypothesis, $J$, for a given number of observed signal events (in the signal region), originating from the true scenario $I$.
We demonstrated that the sphaleron hypothesis can be discriminated from various BH scenarios with a small number of events. Separation between BH scenarios with different minimal BH masses is also possible with a reasonable number of events collected at LHC Run-2, -3 and HL-LHC. The discrimination among the BH scenarios with the same minimal BH mass but a different number of extra dimensions is more challenging and requires a larger number of signal events, which may be collected at future high-energy colliders.

Finally, we emphasise that the work presented in this paper should be considered a proof-of-concept study, as it entirely relies on MC simulations.
In the actual experimental situation, one must validate and correct the MC modelling with the real data before using it for model discrimination.
Such a process is non-trivial and interesting in its own right.
We, however, postpone this subject for future work.

\section*{Acknowledgements}
The research of A.S.G, F.K., A.L., R.M., K.S.\ and T.S\ leading to these results has received 
funding from the Norwegian Financial Mechanism for years 2014-2021, 
grant no DEC-2019/34/H/ST2/00707. 
T.S. acknowledges support from the Research Council of Norway, grant no 314472, which is supporting the research of A.S.G and I.S.
K.S. is supported by the National Science Centre, Poland, under research grant 2017/26/E/ST2/00135.
R.M. is supported by the National Science Centre, Poland, under PRELUDIUM research grant no 2021/41/N/ST2/00972 and under SONATA BIS grant no 2020/38/E/ST2/00243.
A.P. acknowledges support by the National Science Foundation under Grant No.\ PHY 2210161.

\bibliographystyle{utphys28mod}
\bibliography{refs}

\newpage


\appendix

\section{Example event images}

We show below in Fig.\ \ref{event_image}, event images 
of various physics scenarios:
SPH\_9, BH\_n4\_M8,
BH\_n2\_M10, BH\_n4\_M10, BH\_n6\_M10
and 
BH\_n4\_M12.
Three randomly-chosen events in the validation dataset
are used 
for each scenario as examples.
The signals in the ECAL, HCAL and tracking system correspond to 
the red, green and blue colour intensities, respectively.
To aid visualisation, 
we map the accumulated energy deposits ($p_T$ of the tracks), $E_i$, in bin $i$
to the colour intensity $I_i = f(E_i)$
with a conversion function 
$f(E_i) = \arctan( \ln ( E_i/20\,{\rm 
GeV}) )/\pi + \tfrac{1}{2} \in [0, 1]$.
The reconstructed jet 
(anti-$k_T$ jet with radius parameter $R= 0.4$)
positions for each event are indicated with orange circles of radius 0.4.

\vspace{5mm}

\begin{figure}[h!]
     \centering
    \caption{Event images of various physics scenarios.
    Three randomly-chosen events are shown for each scenario.
    The red, green and blue colour intensities correspond to the signals in the ECAL, HCAL and tracking system, respectively.     
    The reconstructed jet positions for each event are indicated with orange circles of radius 0.4.}
    \label{event_image}
     \includegraphics[width=0.30\textwidth]{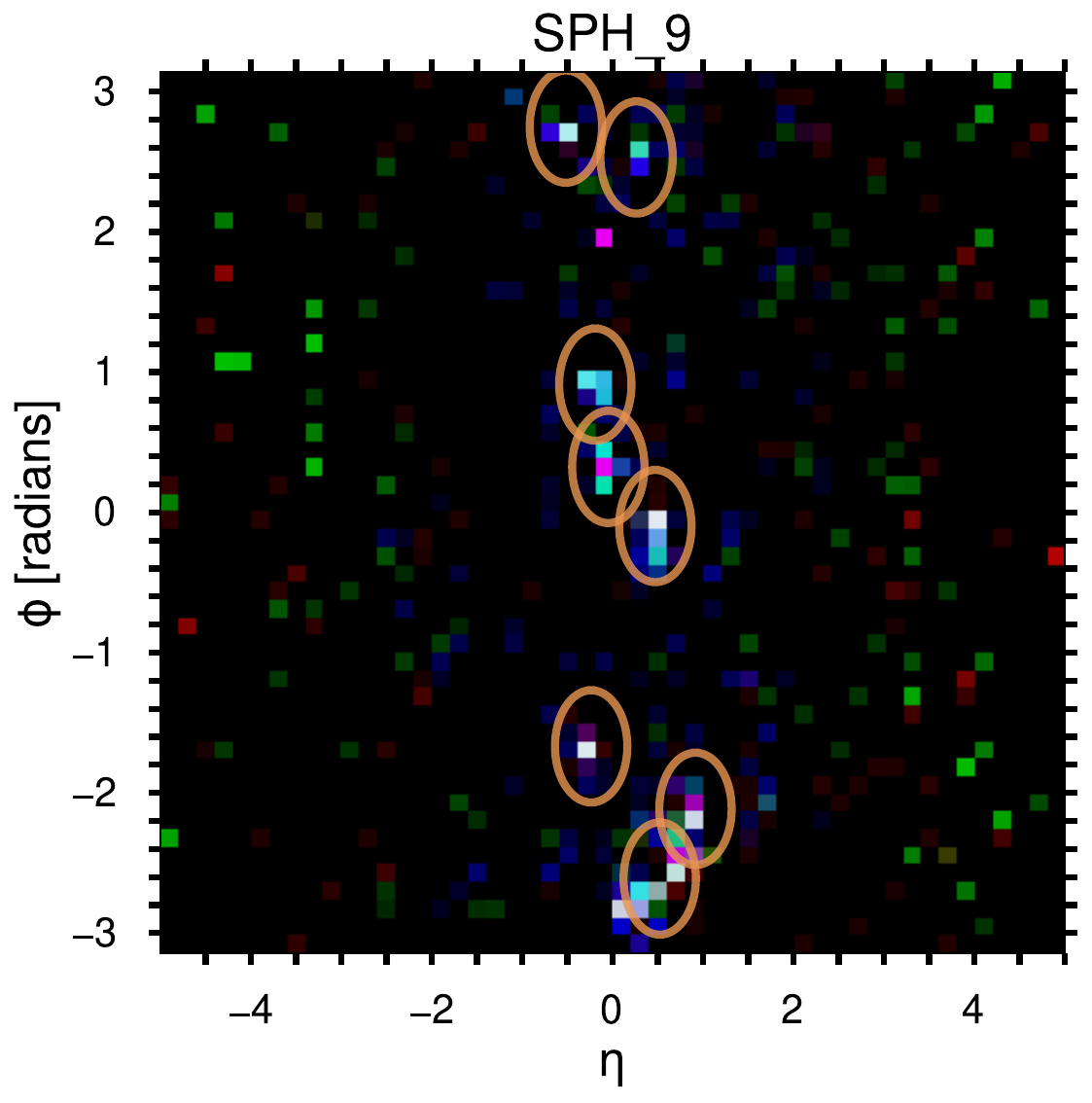}
     \includegraphics[width=0.30\textwidth]{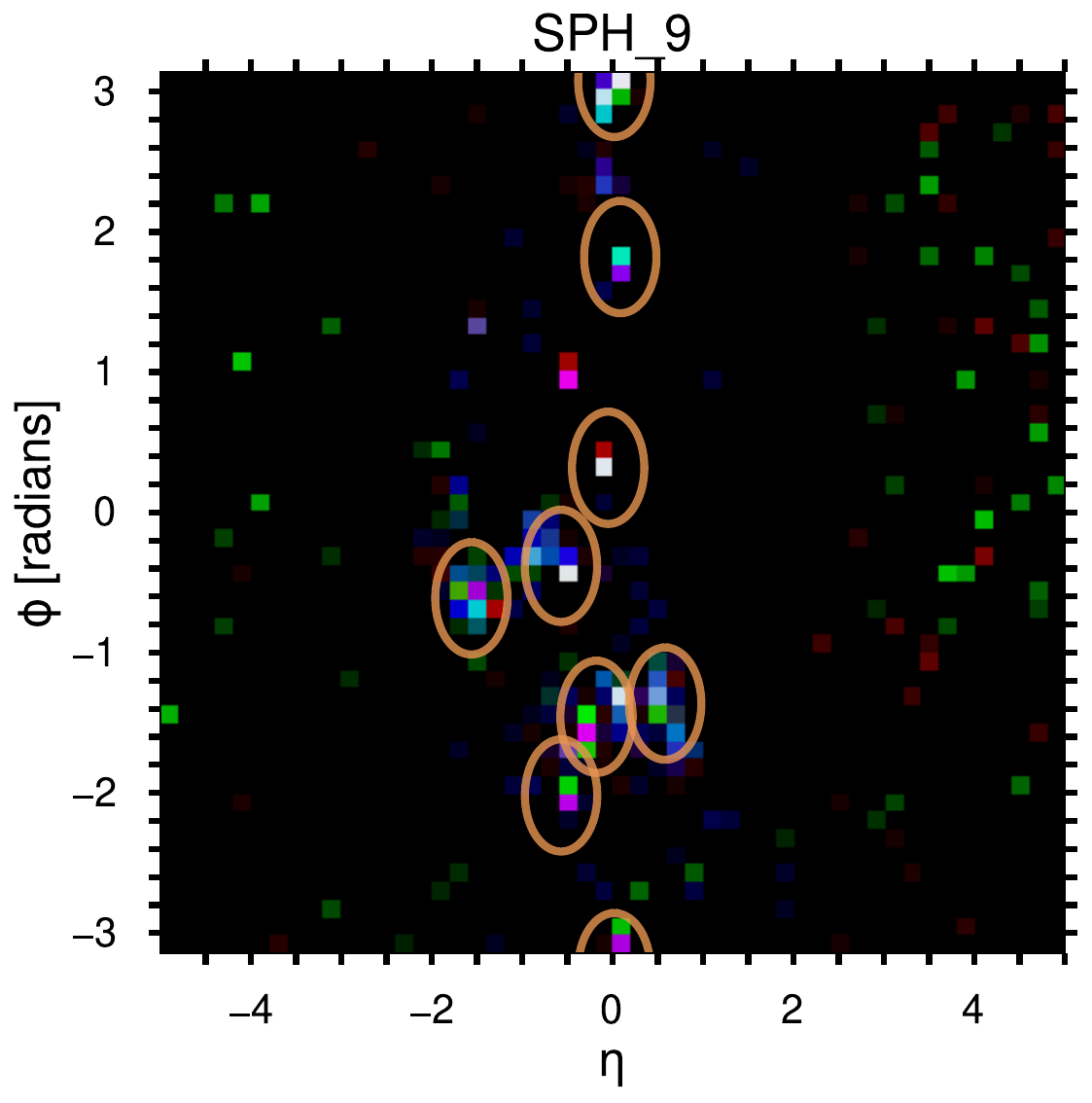}
     \includegraphics[width=0.30\textwidth]{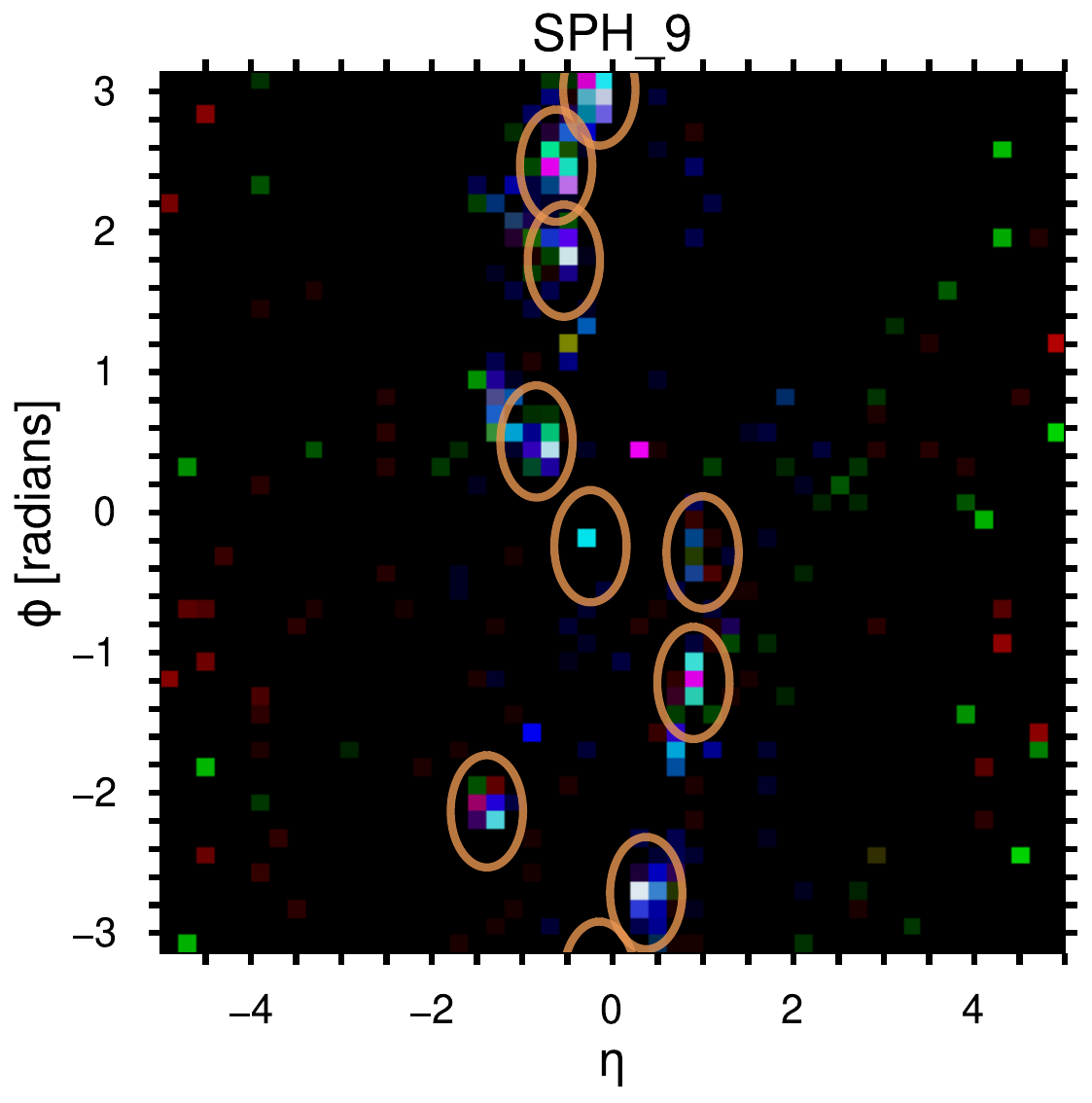}  
     \includegraphics[width=0.30\textwidth]{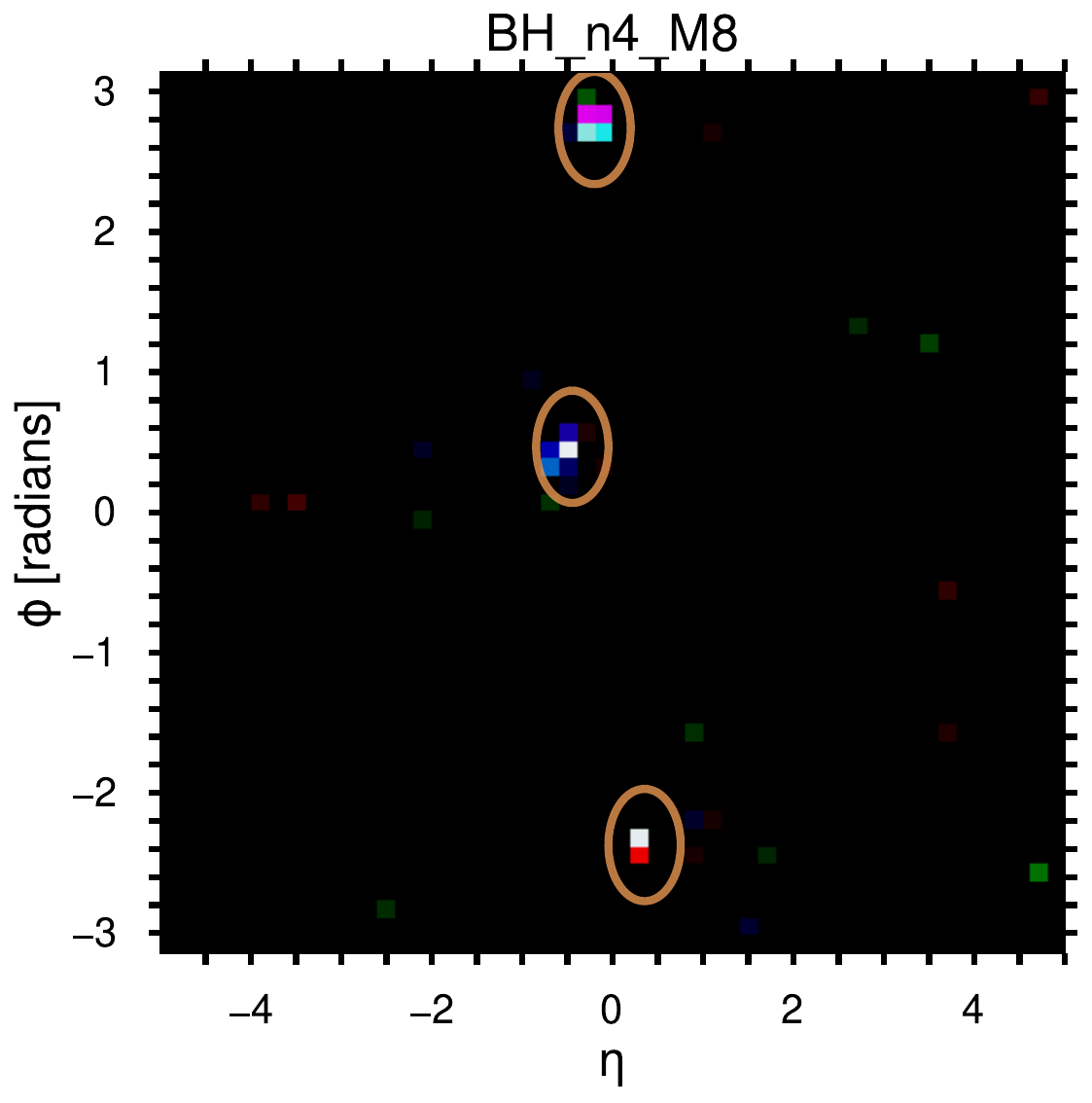}
     \includegraphics[width=0.30\textwidth]{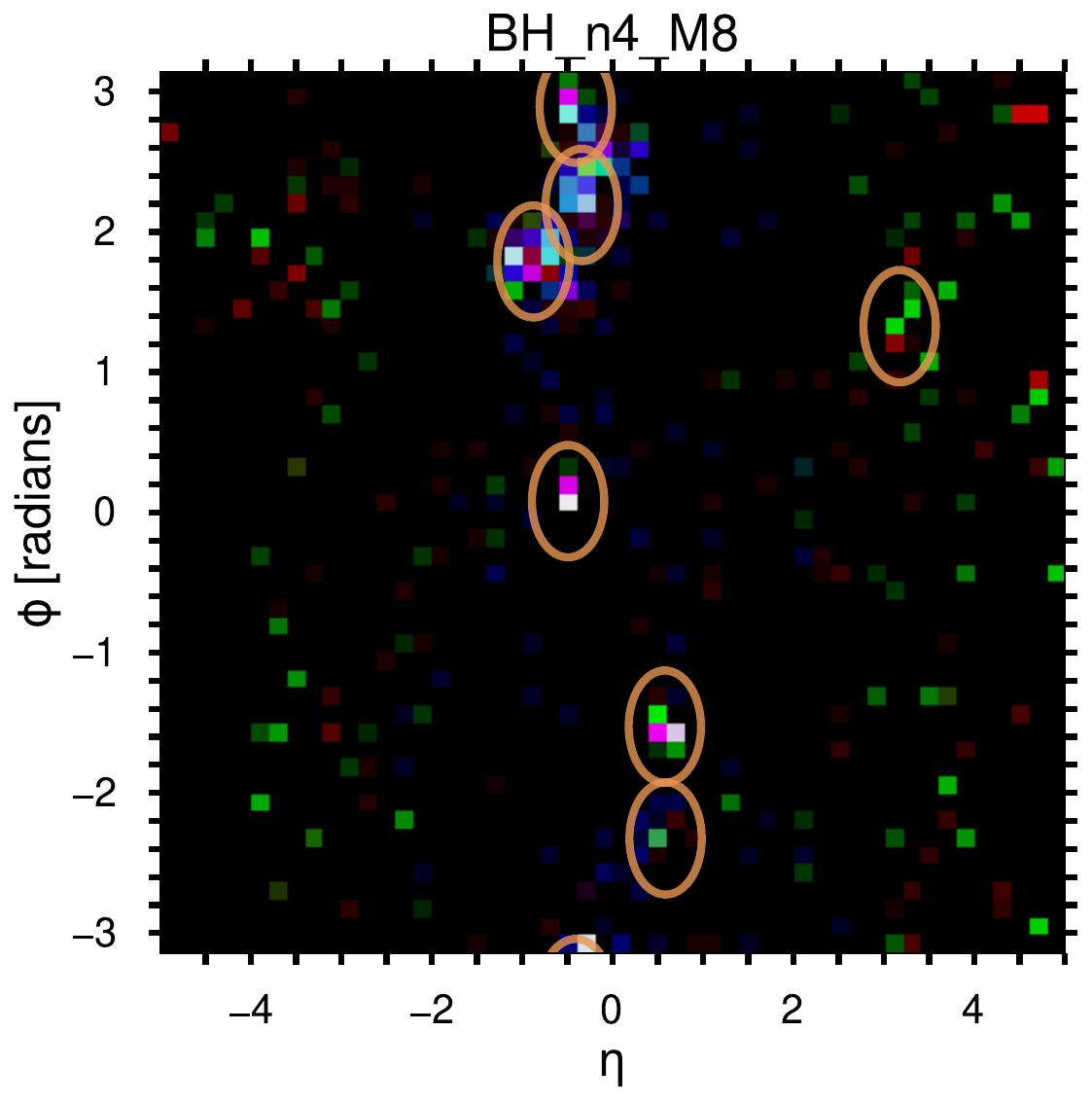}
     \includegraphics[width=0.30\textwidth]{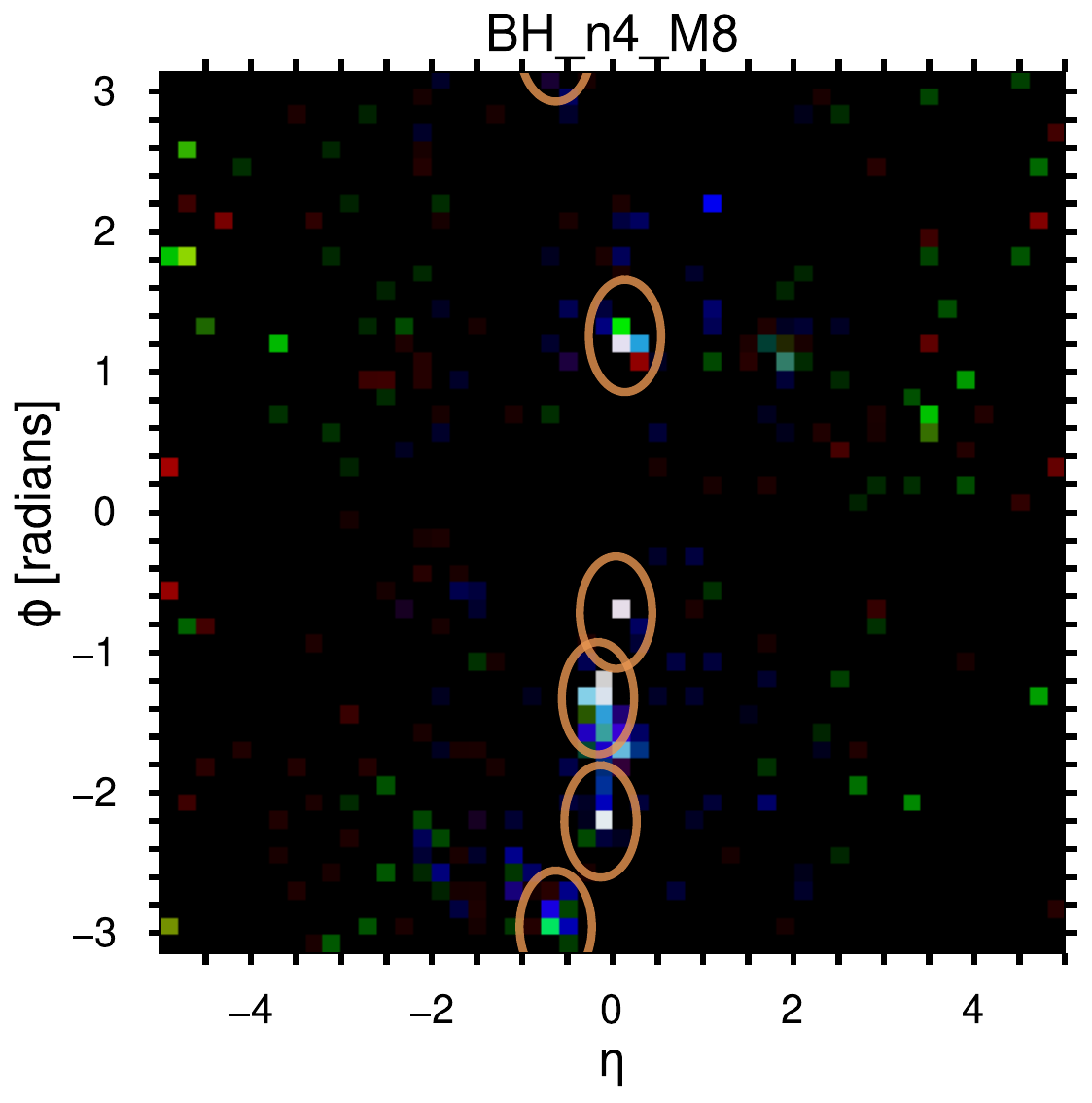}
     \includegraphics[width=0.30\textwidth]{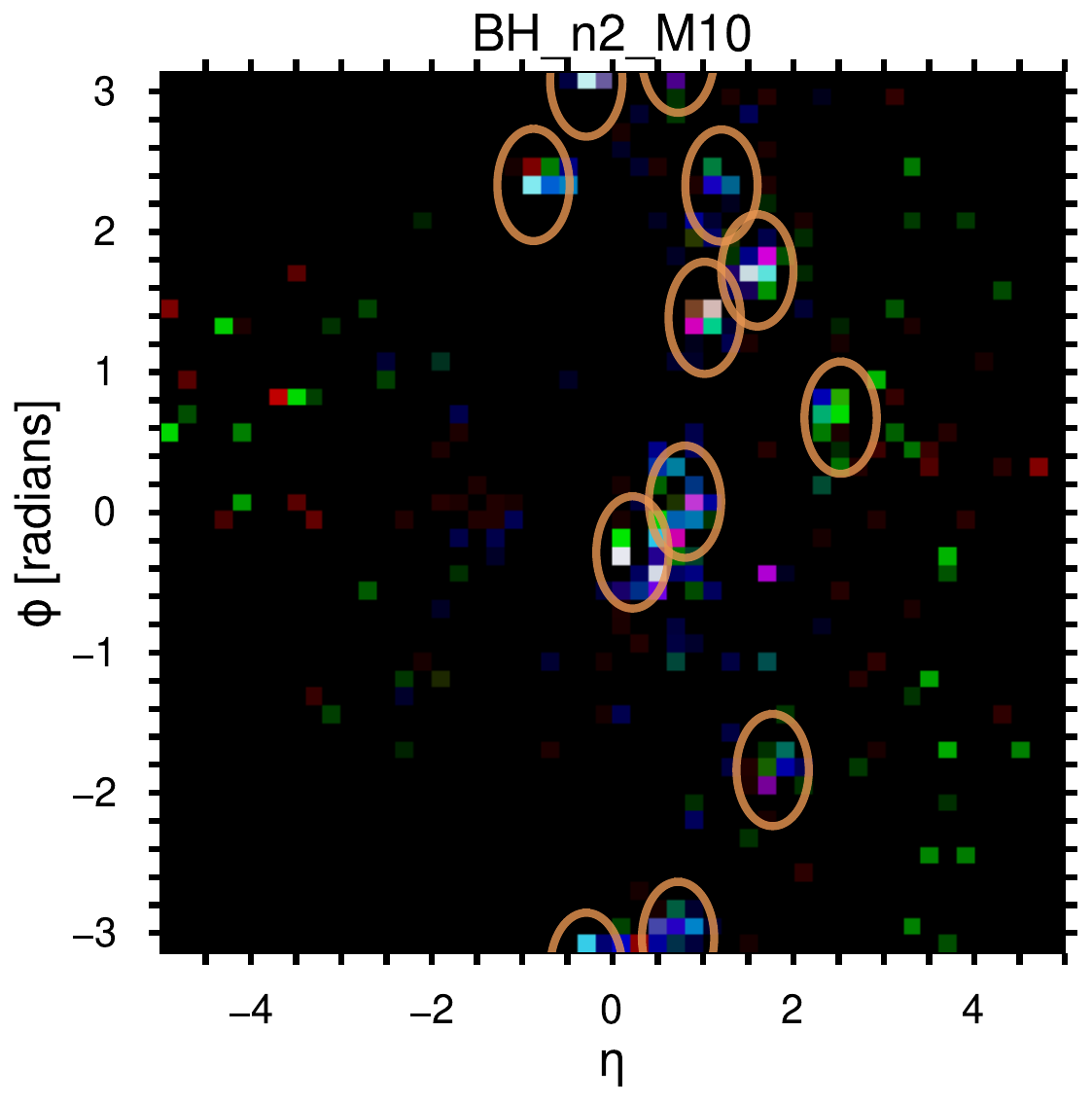}
     \includegraphics[width=0.30\textwidth]{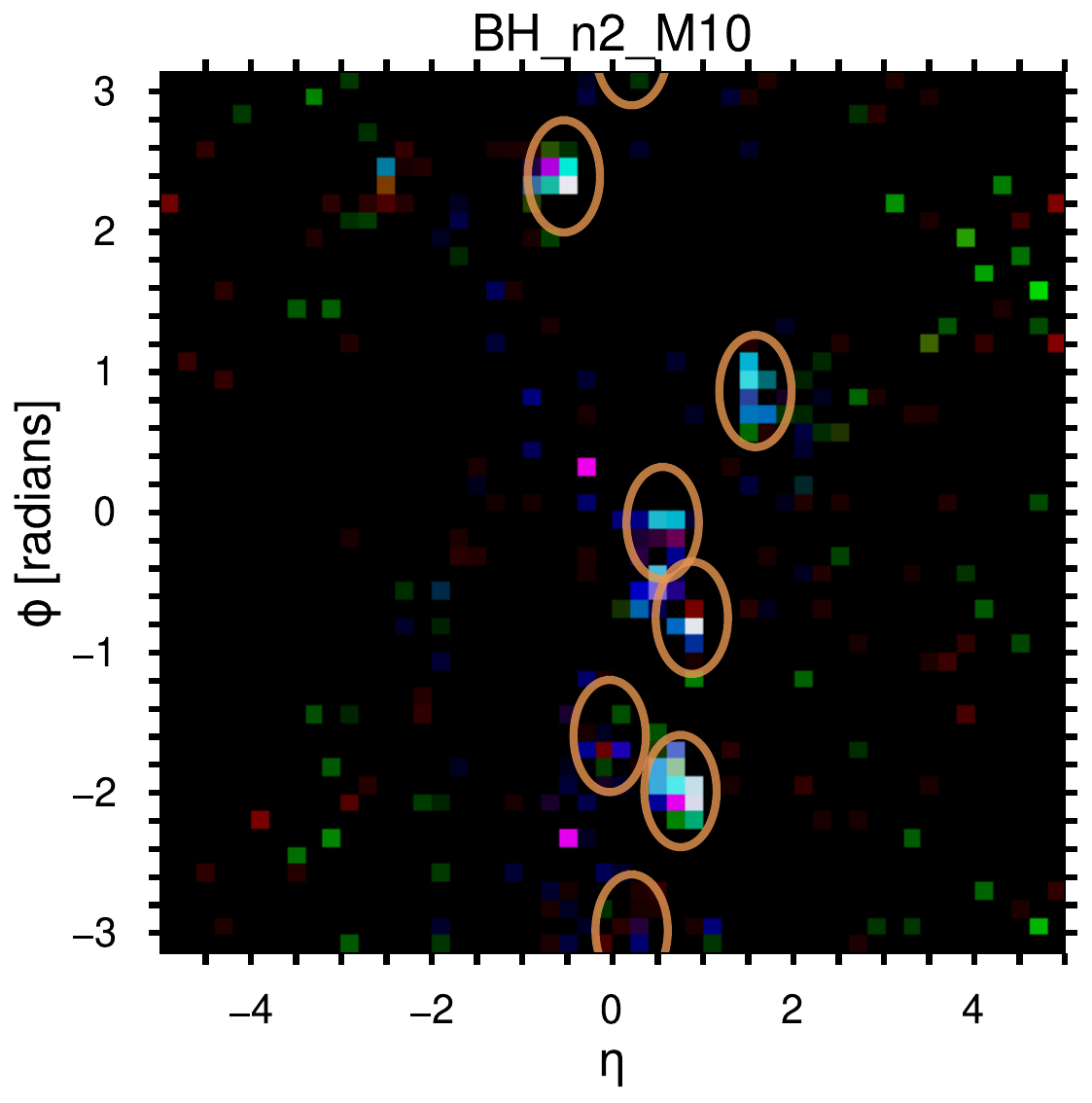}
     \includegraphics[width=0.30\textwidth]{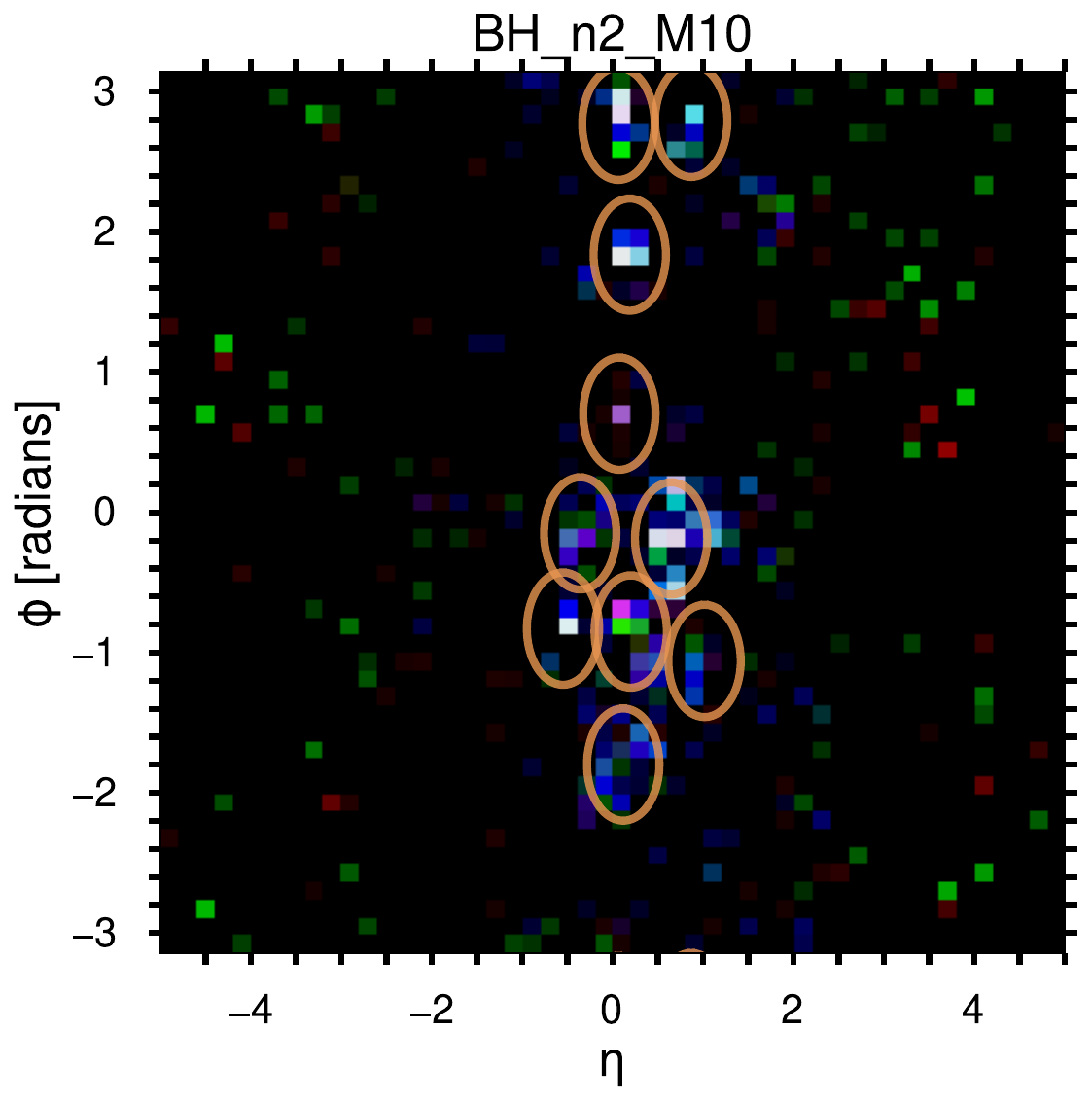} 
\end{figure}

\begin{figure}[h!]
     \centering
     \includegraphics[width=0.30\textwidth]{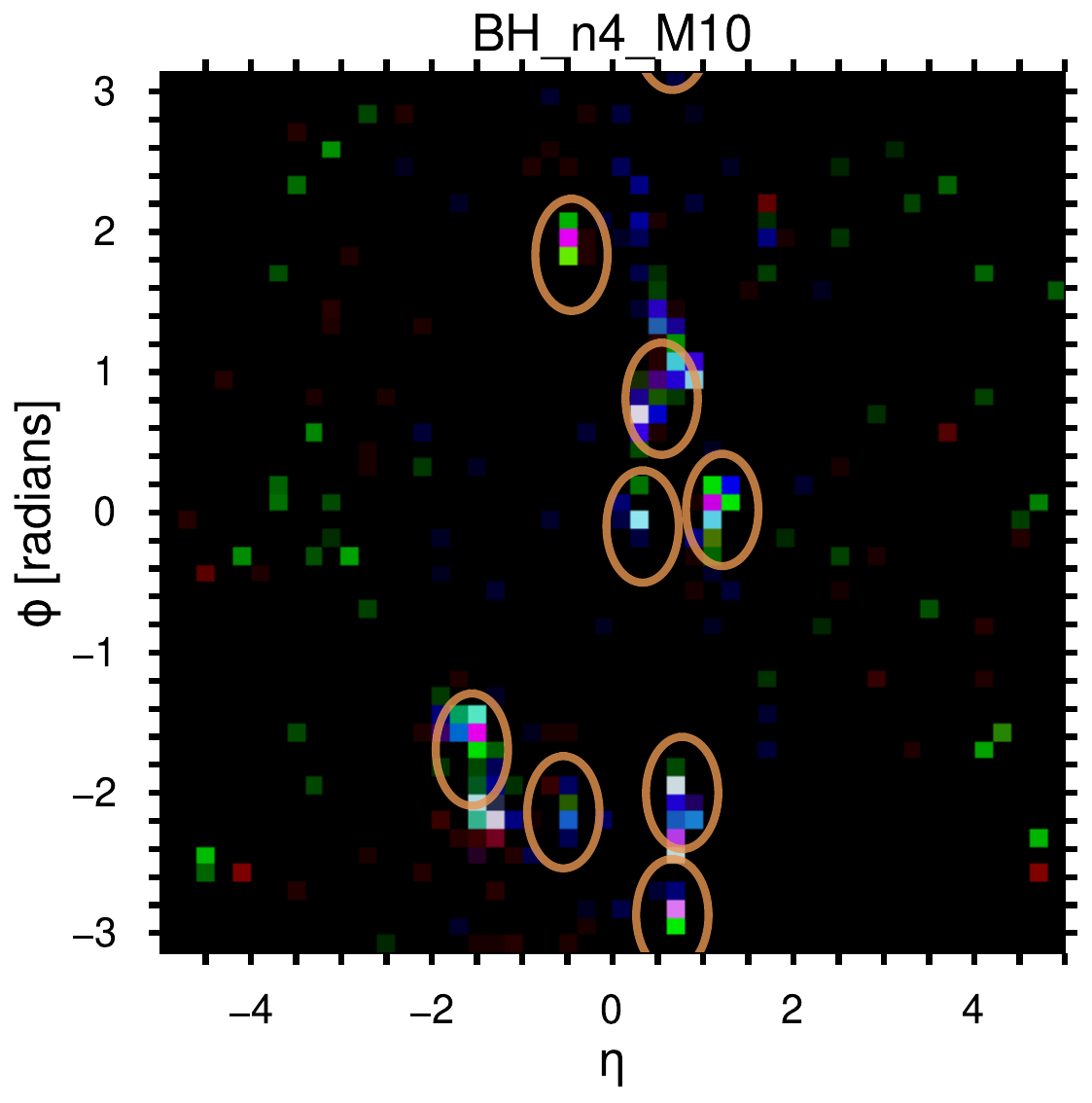}
     \includegraphics[width=0.30\textwidth]{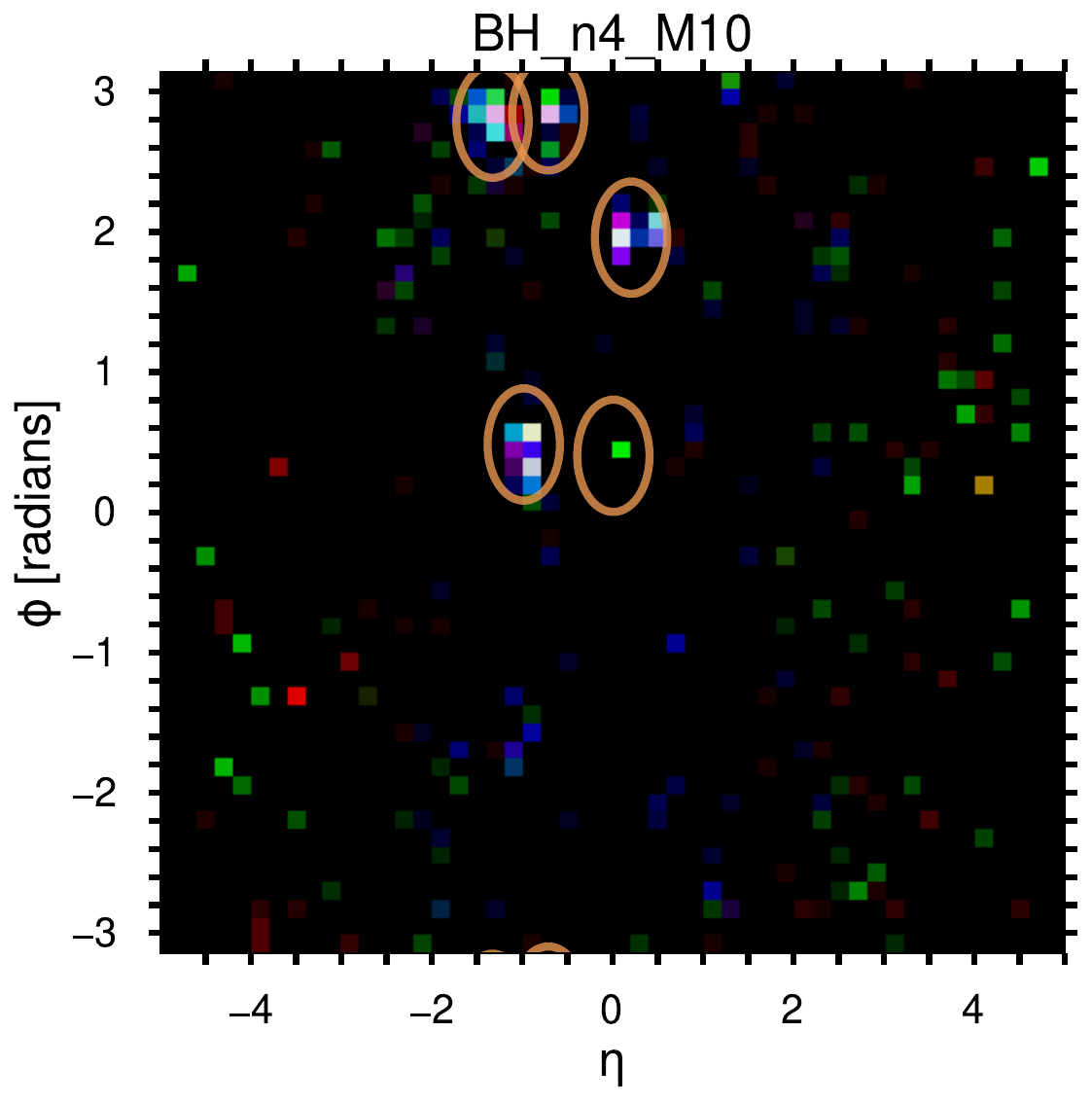}
     \includegraphics[width=0.30\textwidth]{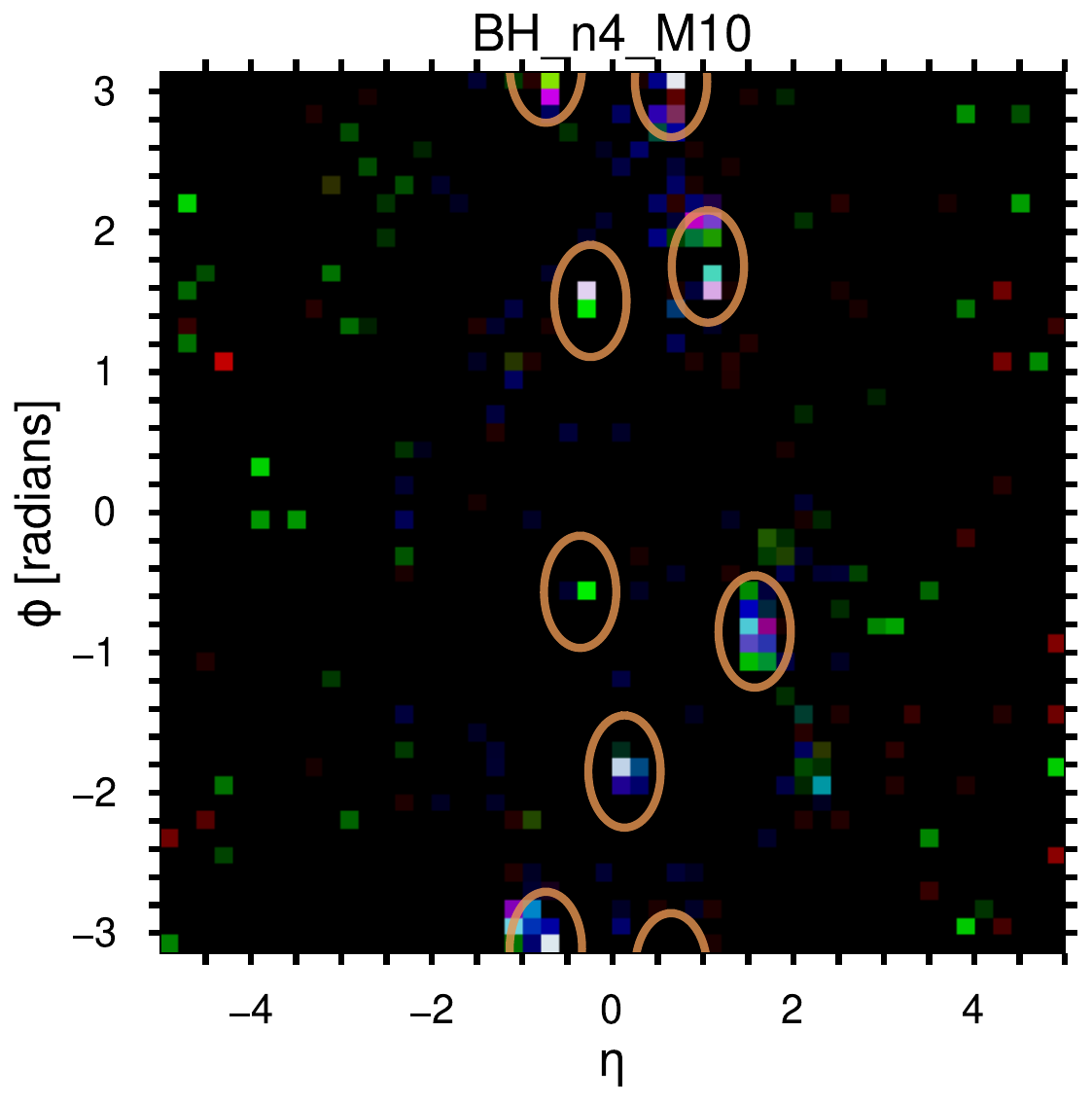}  
     \includegraphics[width=0.30\textwidth]{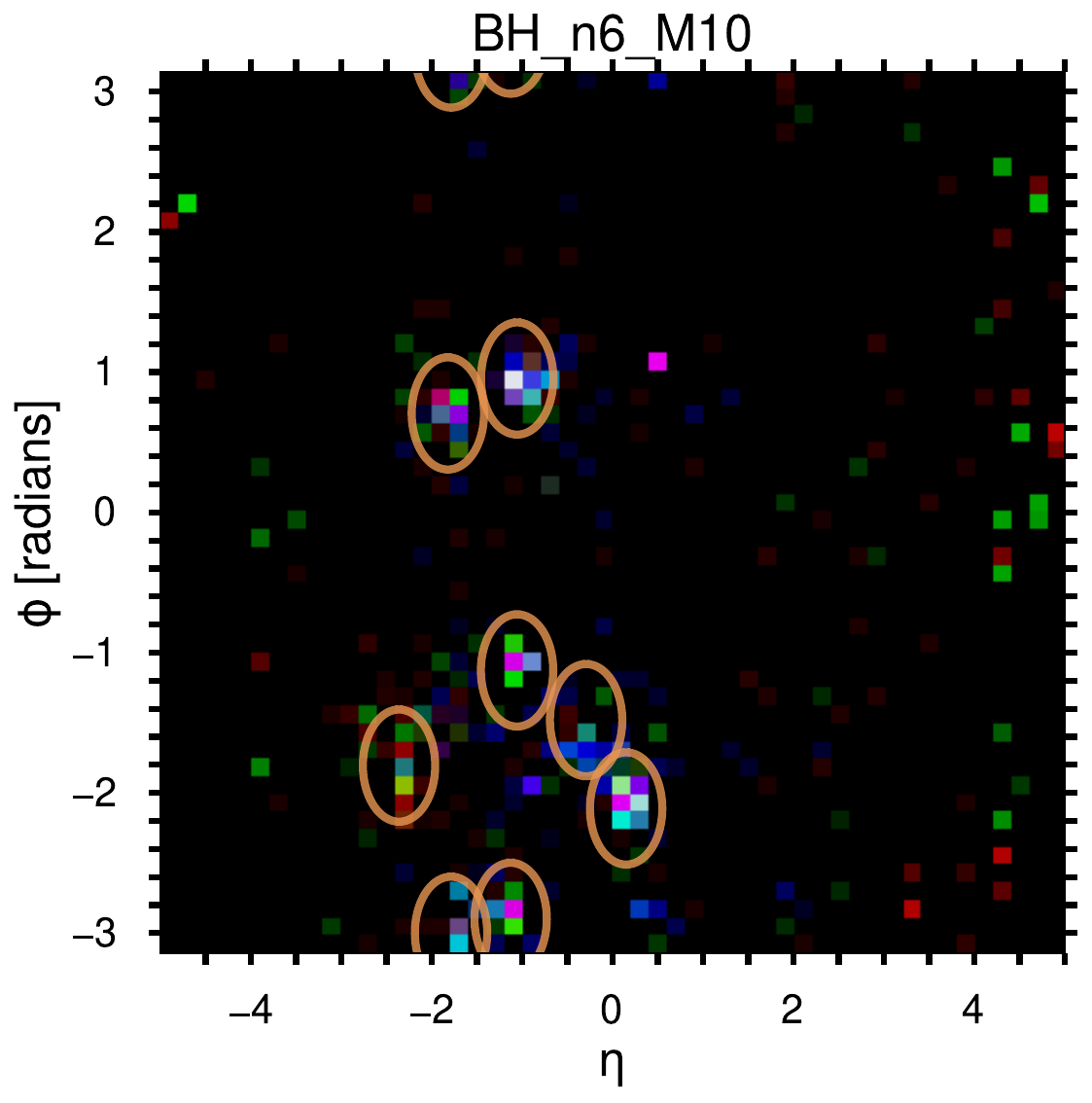}
     \includegraphics[width=0.30\textwidth]{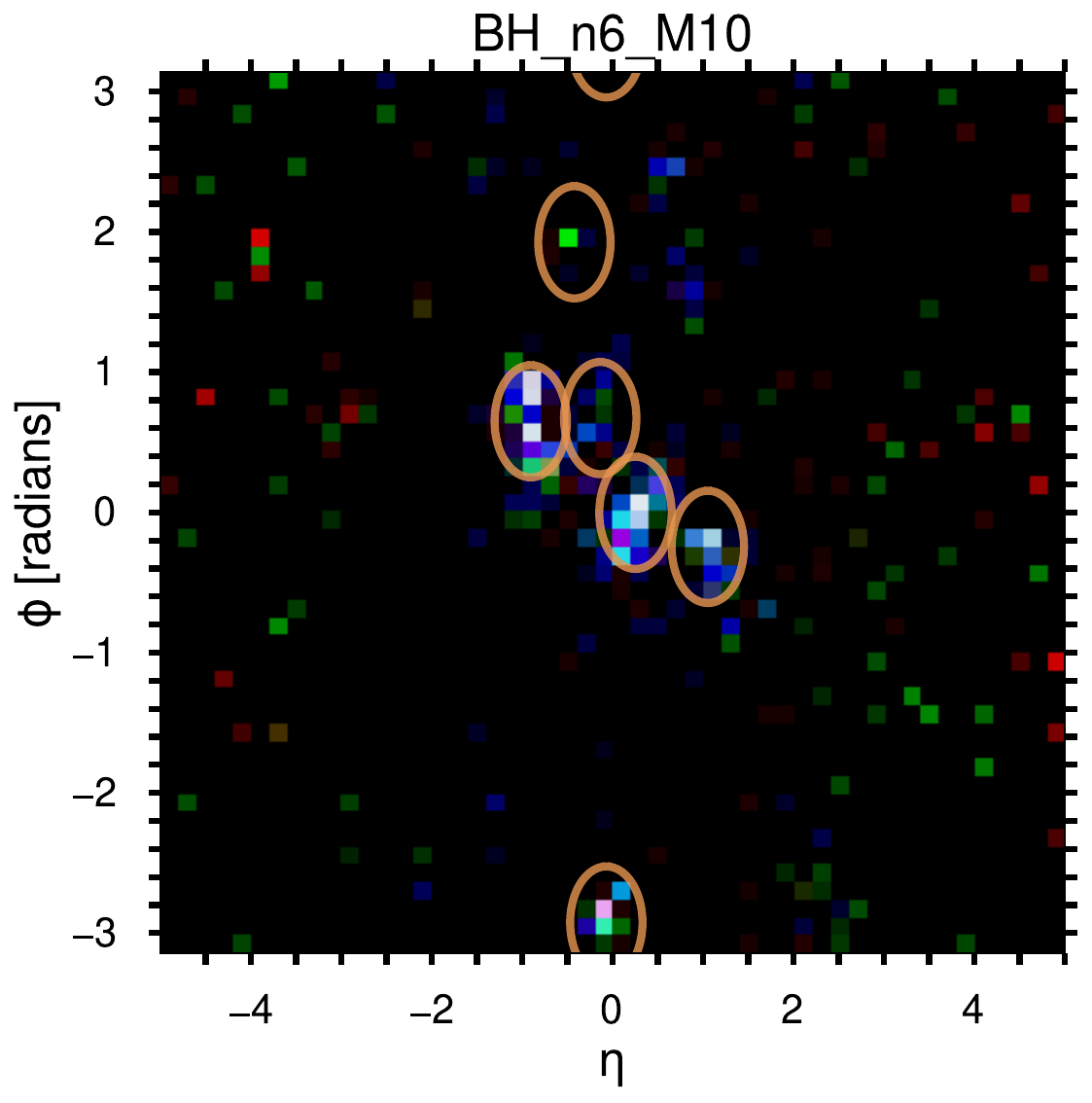}
     \includegraphics[width=0.30\textwidth]{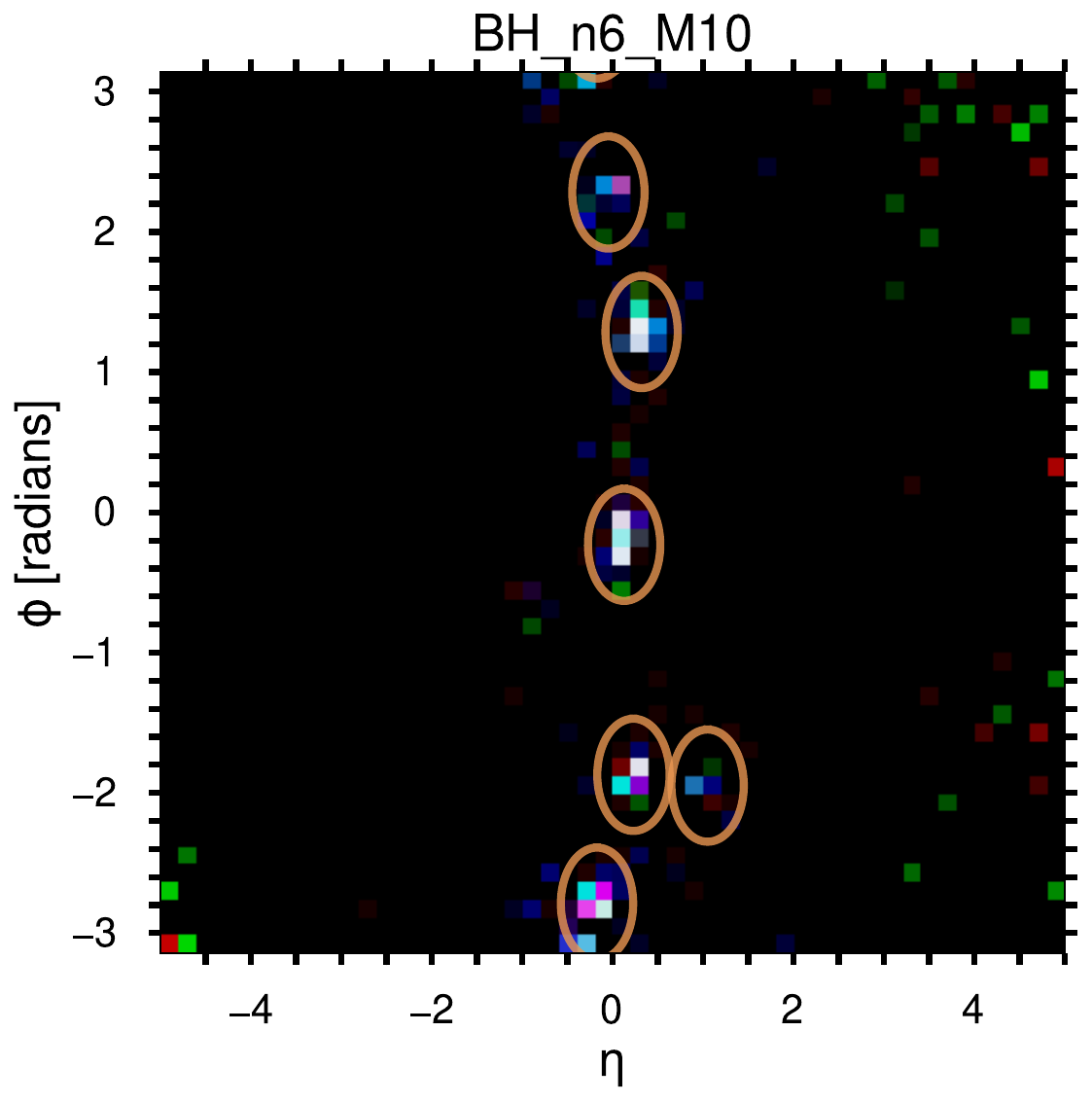} 
     \includegraphics[width=0.30\textwidth]{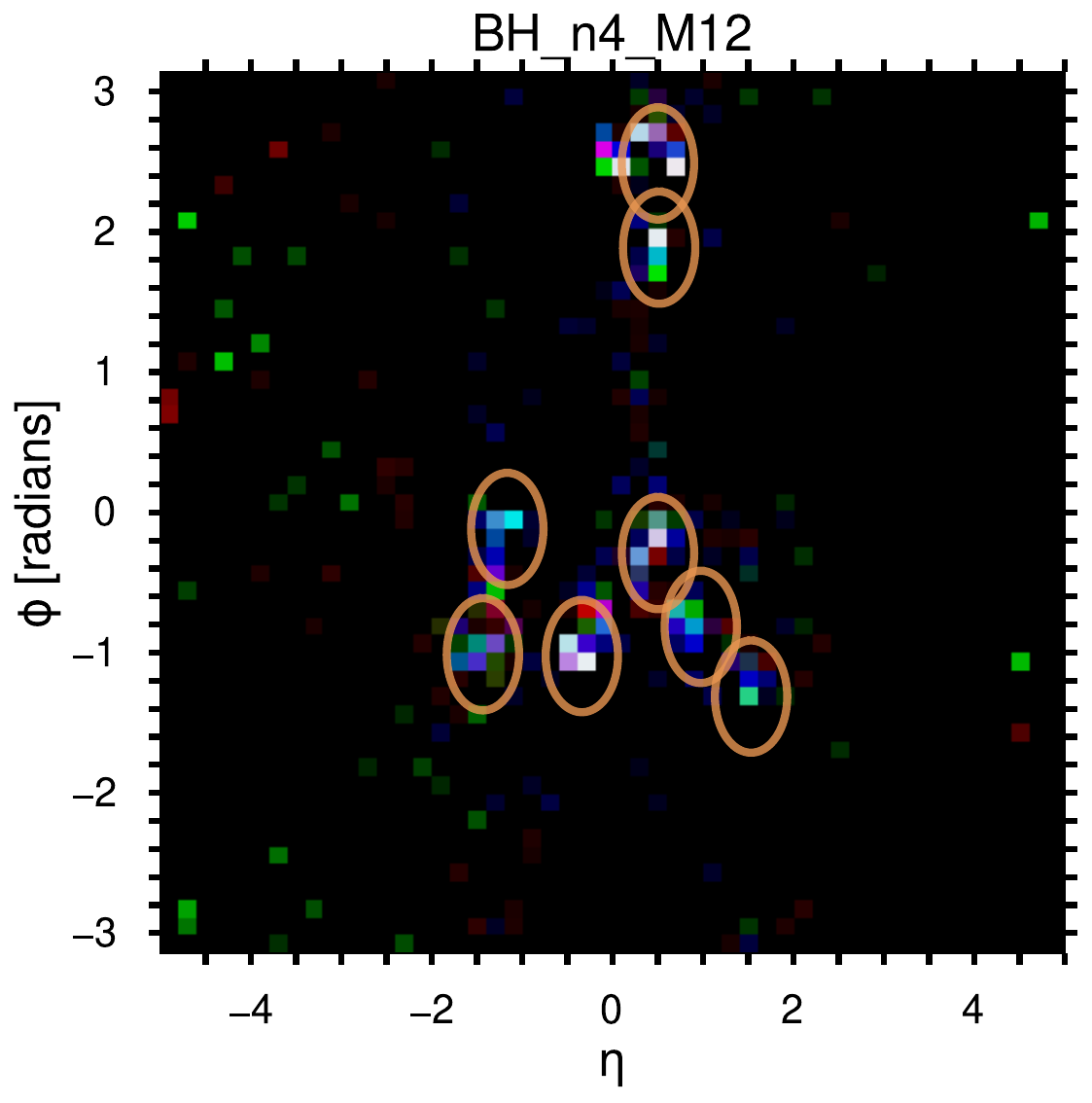}
     \includegraphics[width=0.30\textwidth]{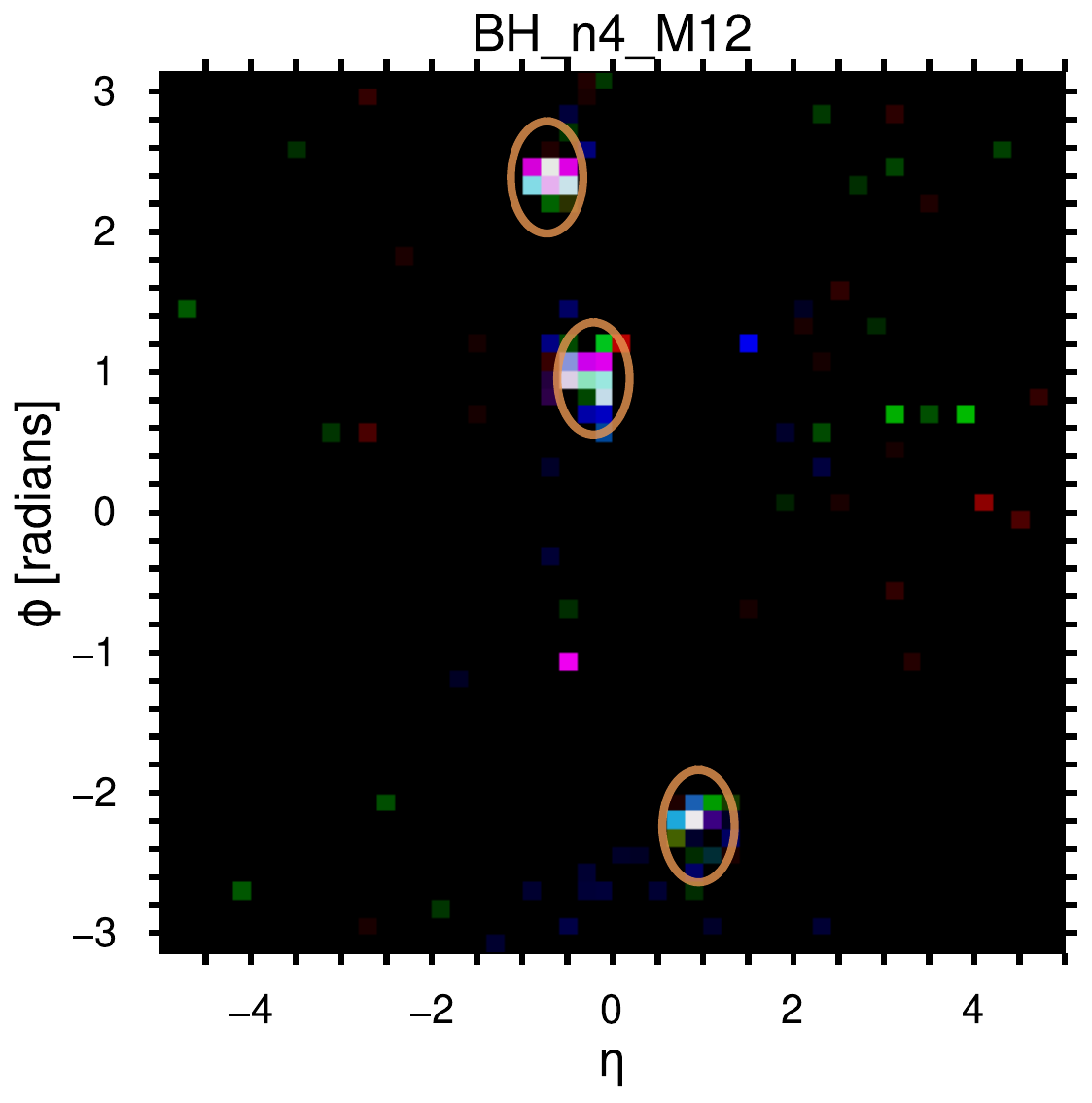}
     \includegraphics[width=0.30\textwidth]{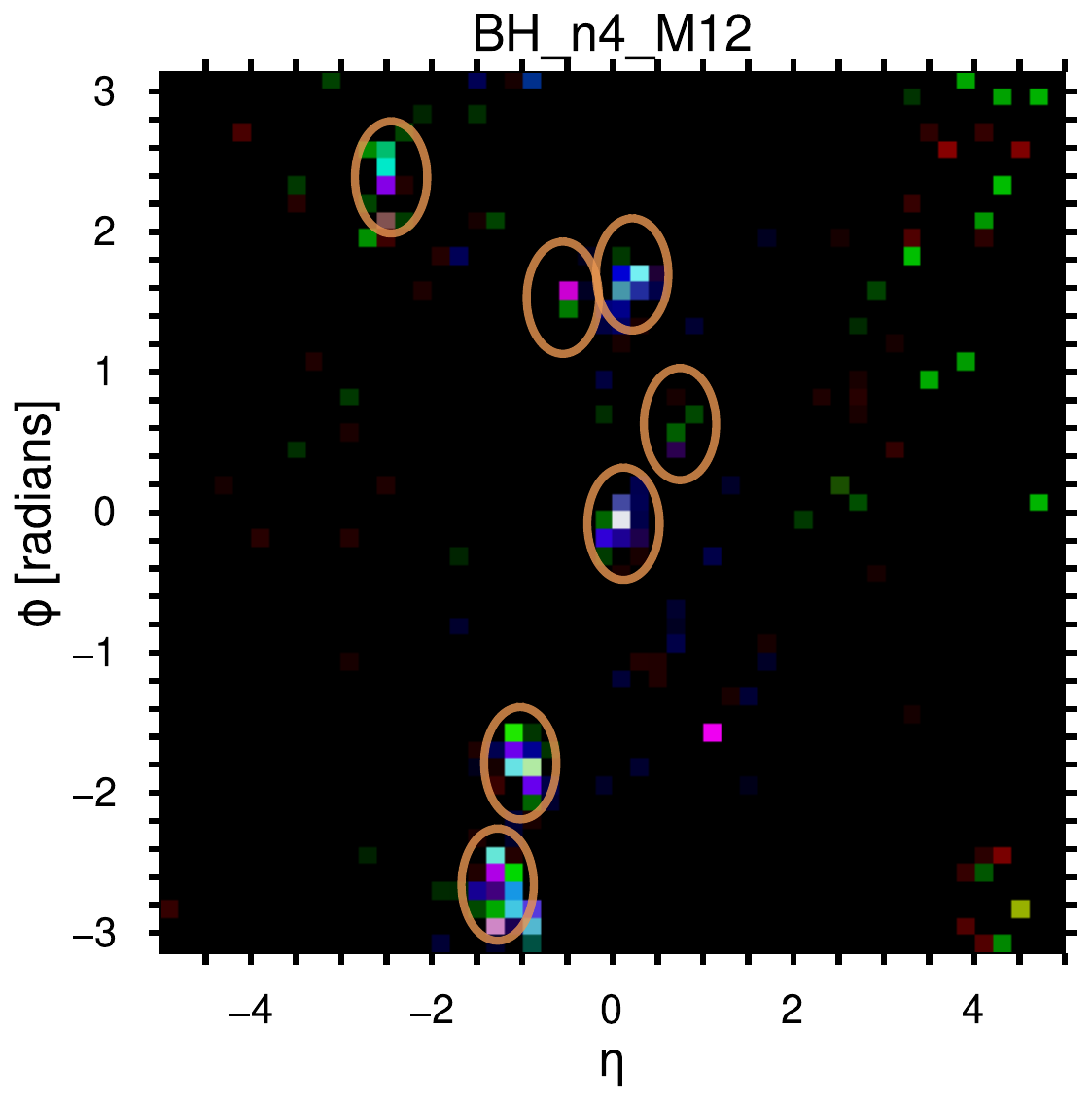}   
\end{figure}


\newpage

\end{document}